\documentclass[amsfonts, amssymb, amsmath, reprint, showkeys, nofootinbib, twoside]{revtex4-1}
\usepackage[english]{babel}
\usepackage[utf8]{inputenc}
\usepackage{amsthm}
\usepackage{mathtools}
\usepackage{xcolor}
\usepackage{color}
\usepackage{graphicx}
\usepackage{adjustbox}
\usepackage{placeins}
\usepackage[T1]{fontenc}
\usepackage{lipsum}
\usepackage{csquotes}
\usepackage[dvipdfmx]{hyperref}
\usepackage{bm}
\usepackage{mathrsfs}
\usepackage{amsfonts}
\begin{document}
\title{
Geometrical Quantum Engine Driven Isothermally by Electrochemical Potentials 
}


\author{Hisao Hayakawa}\email[e-mail address]{hisao@yukawa.kyoto-u.ac.jp}
\affiliation{Yukawa Institute for Theoretical Physics, Kyoto University, Kitashirakawa-oiwake cho, Sakyo-ku, Kyoto 606-8502, Japan}

\author{Ville M.\,M.\,Paasonen}
\altaffiliation[Present Address: ]{Department of Physics and Astronomy, University of Turku, FI-20014 Turun yliopisto, Finland}
\affiliation{Yukawa Institute for Theoretical Physics, Kyoto University, Kitashirakawa-oiwake cho, Sakyo-ku, Kyoto 606-8502, Japan}

\author{Ryosuke Yoshii}
\affiliation{Center for Liberal Arts and Sciences, Sanyo-Onoda City University,
Yamaguchi 756-0884, Japan}
\date{\today}

\begin{abstract}

We propose a geometrical engine undergoing an adiabatic (Thouless) pumping process for a small system connected to external isothermal reservoirs  with the control of electrochemical potentials of the reservoirs and one parameter in the system Hamiltonian.
Thanks to the geometrical nature of this process, the entropy production is characterized by the geometric metric tensor which is connected to the Fisher information and the Hessian of the density matrix in a nonequilibrium steady state.
The existence of an inequality between the thermodynamic length and entropy production is established.
We also establish that the work done on this system is characterized by a vector potential and is equivalent to the thermodynamic flux. 
To characterize the engine, we the introduce effective efficiency as the relation between the work and entropy production.
Through the theoretical analysis of the quantum master equation for the Anderson model of a quantum dot within the wide-band approximation, we illustrate the explicit values of the work, thermodynamic length, and effective efficiency of the engine as functions of the phase difference of the externally controlled electrochemical potentials.

\end{abstract}
\maketitle

\section{INTRODUCTION}\label{sec:intro} 


 Since the first proposal of the adiabatic geometrical pumping by Thouless~\cite{thouless1,thouless2}, it has been recognized that a current can flow without any averaged bias if there is  
a Berry-phase-like variable or the Berry-Sinitsyn-Nemenman (BSN) curvature in the space of the modulation parameters \cite{thouless1,thouless2,berry,xiao,sinitsyn1,sinitsyn2}.
This phenomenon, known as Thouless pumping or geometrical pumping, has been observed experimentally in various processes such as charge transport \cite{ex-ch1,ex-ch2,ex-ch2.5,ex-ch3,ex-ch4,ex-ch5,ex-thou1,ex-thou2} and spin pumping~\cite{ex-spin1}.
There are various theoretical papers on geometrical pumping processes based on scattering theories~\cite{brouwer,s-th1,s-th2,s-th3,s-th-ch1,s-th-ch2,s-th-ch3,s-th-spin1}, classical master equations~\cite{parrondo,usmani,astumian1,astumian2,rahav,chernyak1,chernyak2,ren,sagawa,ville2021} and quantum master equations \cite{qme1,qme2,qme-spin1,qme-spin2,yuge1,yuge2}. 
The extended fluctuation theorem for geometrical pumping processes has also been studied \cite{watanabe,Hino-Hayakawa,Takahashi20JSP}. 

The phenomenon of geometrical pumping can be utilized in nanoscale thermodynamics.
The geometrical framework has already been used in finite-time thermodynamics \cite{finite}, 
in which the thermodynamic length plays a key role.
The thermodynamic length was originally introduced for macroscopic systems \cite{Weinhold,Ruppeiner,Salamon,Schlogl,geo-rev}, and 
it has since been applied to a wide range of thermodynamical systems such as a classical nanoscale system \cite{Crooks}, a closed quantum system \cite{Deffner} and an open quantum system \cite{Scandi}.
These concepts are important, in particular, for quantum thermodynamics or thermodynamics for nanoscale machines~\cite{Haengi09,Gemmer09,Horodecki13}.
The current status of the subject is presented in a recent review~\cite{Wang2021}.
Thus, the geometrical formulation is of fundamental importance in understanding and controlling nanoscale machines. 

There are several attempts to formulate the geometrical thermodynamics for a microscopic heat engine in the adiabatic regime~\cite{Brandner-Saito,engine,Hino2021,Bhandari20,Abiuso20,Alonso21,Wang2021b}.
In particular, it is remarkable that Ref.~\cite{Hino2021} has formulated the geometrical theory of the heat engine induced by the BSN curvature without average temperature difference between the two reservoirs.
Nevertheless, Ref.~\cite{Hino2021} has, at least, two shortcomings.
First, its analysis is only applicable to systems obeying a classical master equation, which means that the quantum effects of the heat engine cannot be discussed.
Second, the implementation of the microscopic heat engine may not be realistic, because the periodic control of the temperatures of the two reservoirs is difficult in practice.
If we replace the temperatures with electrochemical potentials, such a system would be easy to realize experimentally.
Thus, in this paper, we control the electrochemical potentials of reservoirs isothermally.
We apply the formulation developed in Ref.~\cite{Hino2021} to the Anderson model, a single quantum dot system described by a quantum master equation within the wide-band approximation~\cite{Yoshii13}.
 
The organization of this paper is as follows. 
In Sec. \ref{sec:general}, we explain the setup and the geometrical formulation for describing the heat engine under an adiabatic pumping process. 
In Sec. \ref{sec:appli}, we apply our formulation to the Anderson model for a quantum dot coupled to two reservoirs within the wide-band approximation.
In Sec.~\ref{sec:discussion} we discuss our results including (i) whether a perfectly periodic engine is possible, (ii) what happens if we control the temperature of a reservoir, and (iii) Cram\'{e}r-Rao bound for this engine~\cite{Cramer46,Rao45,Ito2020}.
Finally,  in Sec. \ref{sec:conclusion} we summarize our results and perspectives. 
In Appendix \ref{sagawa_KL_divergence}, we prove the positivity of the quantum relative entropy.
In Appendix \ref{housekeeping} we discuss the housekeeping entropy.
In Appendix \ref{app:slow-driving}, we describe some general properties of the quantum master equation such as the perturbation method with a slowly modulated parameter and a mathematical description of the pseudo-inverse of the transition matrix.
In Appendix \ref{K-matrix} we derive the transition matrix of a quantum master equation for the Anderson model.
In Appendix \ref{sec:K^+} we present the explicit form of the pseudo-inverse matrix for the Anderson model.
In Appendix \ref{app:detailed_Anderson}, we summarize some detailed properties on the Anderson model.
In Appendix \ref{app:perturbed_Anderson}, we present the detailed calculation of the perturbation method for the Anderson model.

\section{GENERAL FRAMEWORK}\label{sec:general}

\subsection{Setup}\label{sec:setup}

In this paper, we consider a small quantum system S such as a quantum dot coupled to two reservoirs L and R under periodic modulation of some system parameters, with the period $\tau_{\mathrm{p}}$. 
Each reservoir $\alpha=$L or R is characterized by the electrochemical potential $\mu^{\alpha}$ and temperature $T$ (or the inverse temperature $\beta:=1/T$).
We adopt the unit that $T$ has the dimension of the energy in this paper.
Although we will specify the system Hamiltonian attached to two reservoirs later, we assume that we can control some parameters in the Hamiltonian through a control parameter $\lambda$. 

\begin{figure}
\centering
\includegraphics[clip,width=8cm]{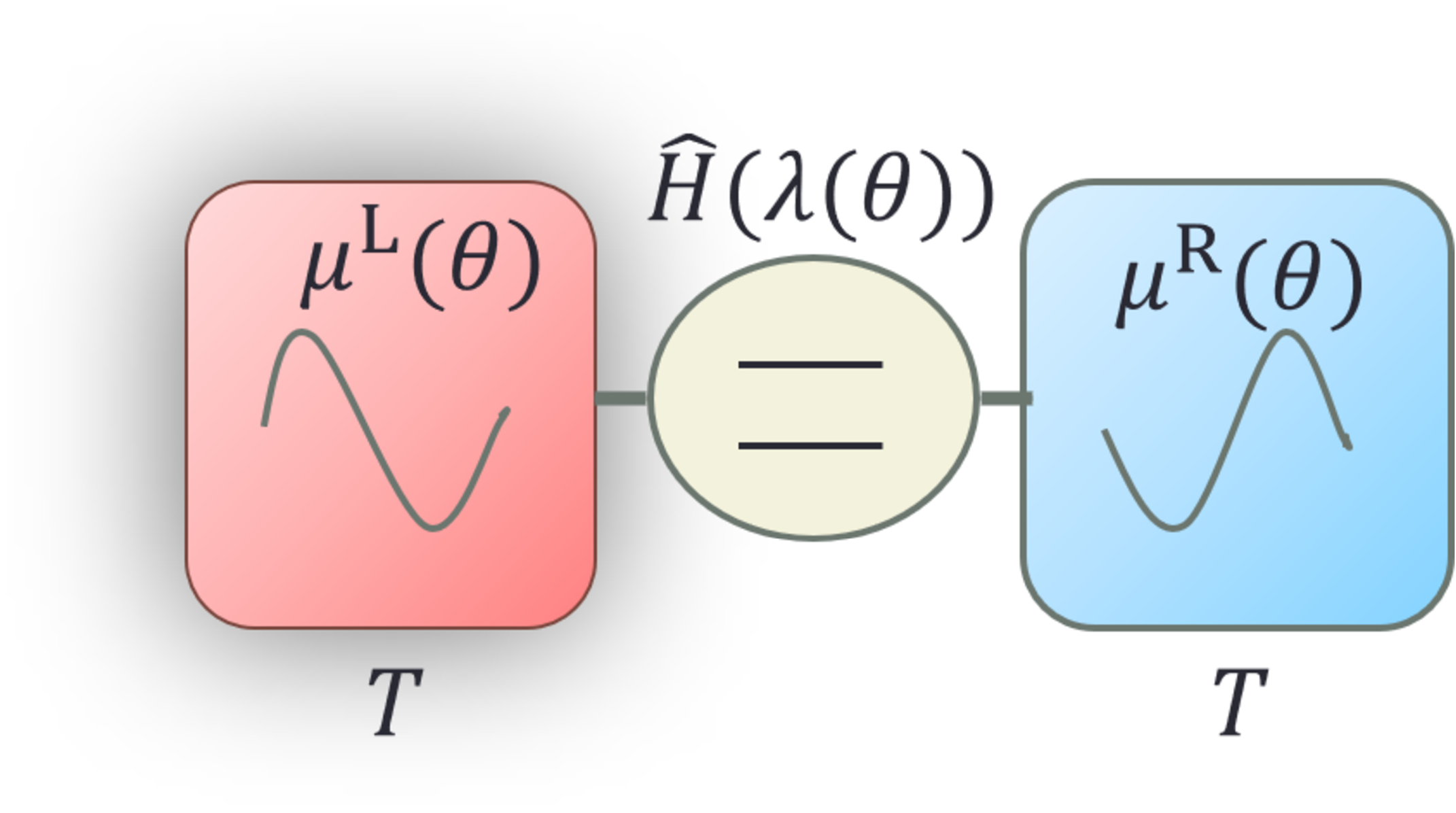}
\caption{
A schematic of the model we consider in this paper, in which $\mu^{\rm L}(\theta)$, $\mu^{\rm R}(\theta)$ and $\hat{H}(\lambda(\theta))$ are modulated by an external agent. }
\label{fig0}
\end{figure}

We illustrate the setup by a schematic picture in Fig.~\ref{fig0} where we modulate $\mu^{\rm L}(\theta)$ and $\mu^{\rm R}(\theta)$ in the reservoirs with different phases. 
The system Hamiltonian $\hat{H}(\lambda(\theta))$ is also modulated by an external agent.


In this paper, we control the set of parameters 
\begin{equation}\label{control_parameters}
\bm{\Lambda} := 
\left(
\lambda,
\frac{\mu^{\mathrm{L}}}{\overline{\mu^{\rm L}}},
\frac{\mu^{\mathrm{R}}}{\overline{\mu^{\rm R}}} 
\right) ,
\end{equation}
where we have introduced
\begin{equation}
\overline{\mu^\alpha}:=\frac{1}{\tau_p}\int_0^{\tau_p}dt\mu^\alpha(t)
\end{equation}
for $\alpha=$L and R. 
When we consider the case without any difference in time-averaged electrochemical potentials, i.e.\, $\overline{\mu^\mathrm{R}}=\overline{\mu^\mathrm{L}}$, 
 we can replace $\overline{\mu^\alpha}$ in Eq.~\eqref{control_parameters} by $\overline{\mu}$ which is independent of $\alpha$.

We assume that the dynamics of S is described by the quantum master equation
\begin{align}\label{master}
    \frac{d}{d\theta} |\hat{\rho}(\theta)\rangle 
    = \epsilon^{-1} \hat{K}(\bm{\Lambda}(\theta)) |\hat{\rho}(\theta)\rangle,
\end{align}
where $\hat{\rho}(\theta)$ is the density matrix of the system S. 
Here, we have introduced the dimensionless time (which is the phase of the
 modulation) $\theta := 2\pi (t-t_{0})/\tau_{\mathrm{p}}$ and the dimensionless operation speed  $\epsilon := 1/(\tau_{\mathrm{p}} \Gamma)$, 
where  $t_{0}$ is the time after which the system reaches a periodic state and $\Gamma$ is the coupling strength or the characteristic transition rate between the system and the reservoirs.
Note that Ref.~\cite{Yoshii22} discussed the relaxation from a nonequilibrium steady state in which the geometric term plays an important role and such a term is negligible after the system reaches a quasi-steady state.
In this sense, we can safely ignore the geometrical term in the density matrix.
We also note that
 the dynamics of physical relevant quantum master equations such as the Lindblad equation are completely positive and trace-preserving (CPTP)~\cite{Lindblad76,Manzano20,Sagawa20}. 
Moreover, we assume that Eq.~\eqref{master} is a Markovian type master equation, though non-Markovianity might be important for periodic modulations of parameters~\cite{Mizuta21}. 
Since we are interested in the dynamics under slow modulations, we also assume that the dynamical effect of $\hat{\rho}(\theta)$ appears through the change of $\bm{\Lambda}(\theta)$.
If the diagonal elements of the density matrix $\hat{\rho}(\theta)$ are decoupled from the off-diagonal elements, the system can be regarded as classical, but if the diagonal elements are mixed with off-diagonal elements, the system is regarded as quantum.

In Eq.~\eqref{master} we have introduced the vector consisting of the elements of the density matrix
\begin{equation}\label{rho_vec}
|\hat{\rho}(\theta)\rangle := 
\begin{pmatrix}
\rho_{11}(\theta)\\
\rho_{12}(\theta)\\
\dots,\\
\rho_{nn}(\theta)\\
\end{pmatrix}
, 
\end{equation}
where $\rho_{ij}(\theta)$ is the $(ij)-$element of $\hat{\rho}$ at $\theta$, which is assumed to be a $n\times n$ matrix.
The density matrix  $\hat{\rho}(\theta)$ satisfies the condition of  probability conservation ${\rm Tr}\hat{\rho}(\theta)=1$.
Thus, the transition matrix $\hat{K}(\bm{\Lambda}(\theta))$ is a superoperator acting on $|\hat{\rho}(\theta)\rangle$, which is expressed as a $n^2\times n^2$ matrix.

We assume that the master equation (\ref{master}) has a unique steady state $|\hat{\rho}^{\mathrm{ss}}(\bm{\Lambda}(\theta)) \rangle$ which satisfies 
\begin{equation}\label{steady_condition}
\hat{K}(\bm{\Lambda}(\theta))|\hat{\rho}^{\mathrm{ss}}(\bm{\Lambda}(\theta)) \rangle = 0.
\end{equation}
Equation \eqref{steady_condition} means that the steady density matrix is equivalent to the right zero eigenstate of $\hat{K}(\bm{\Lambda}(\theta))$.  
Since the system is coupled to two reservoirs having different  electrochemical potentials,  $|\hat{\rho}^{\mathrm{ss}}(\bm{\Lambda}(\theta)) \rangle$ is a nonequilibrium steady state.

The matrix $\hat{K}(\bm{\Lambda}(\theta))$ consists of the transition matrices between the reservoir $\alpha$ to the system as
$\hat{K}(\bm{\Lambda}(\theta))
= \sum_{\alpha=\mathrm{L},\mathrm{R}} \hat{K}_{\alpha}(\bm{\Lambda}(\theta))$.
From probability conservation, the element $k^\alpha_{ij}(\bm{\Lambda}(\theta))$ of $\hat{K}_\alpha(\bm{\Lambda}(\theta))$ satisfies $\sum_{\alpha={\rm L,R}}\sum_i {k}^\alpha_{ij}(\bm{\Lambda}(\theta))= 0$, 
where $k_{ij}^\alpha(\bm{\Lambda}(\theta))$ expresses the transition rate of $j$ to $i$ due to interaction with the reservoir $\alpha$ at $\theta$.
We assume that the $\theta$-dependence of ${k}_{ij}^\alpha(\bm{\Lambda}(\theta))$ only appears through the control parameters $\bm{\Lambda}(\theta)$.
We also assume that $k_{ij}^\alpha(\bm{\Lambda}(\theta))$ satisfies the detailed balance relation: 
\begin{equation}\label{DB}
\sum_{\alpha={\rm L,R}}k_{ij}^\alpha \rho_j^{\rm ss} 
= \sum_{\alpha={\rm L,R}} k_{ji}^\alpha \rho_i^{\rm ss} ,
\end{equation}
where $\rho_i^{\rm ss}$ is the $i-$th component of $|\hat{\rho}^{\rm ss}\rangle$.

\subsection{Thermodynamic Quantities}\label{sec:thermodynamic}

The second law of thermodynamics governs the performance of an engine.
It is well-known that the {\it quantum} relative entropy 
\begin{equation}\label{QKL}
S^{\rm HS}(\hat{\rho}||\hat{\rho}^{\rm ss}):=
{\rm Tr}[\hat{\rho}(\ln\hat{\rho}-\ln\hat{\rho}^\mathrm{ss})],
\end{equation}
 with density matrices $\hat{\rho}$ and $\hat{\rho}^\mathrm{ss}$ satisfying ${\rm Tr}\hat{\rho}={\rm Tr}\hat{\rho}^\mathrm{ss}=1$
is positive semidefinite (see Appendix \ref{sagawa_KL_divergence}) ~\cite{Sagawa20,Petz86,Petz03,Ruskai02,Hiai11,Hatano-Sasa}:
\begin{equation}\label{monotonicity_main}
S^{\rm HS}(\hat{\rho}||\hat{\rho}^{\rm ss})\ge 0 .
\end{equation}
This $S^{\rm HS}$ is regarded as the quantum version of Hatano-Sasa entropy~\cite{Hatano-Sasa}.
We can rewrite $S^{\rm HS}$ as
\begin{align}\label{HS}
    S^{\mathrm{HS}}(\hat{\rho}||\hat{\rho}^{\rm ss}) 
    :=-{S}^{\rm vN}(\theta) + S^{\mathrm{ex}}(\theta),
\end{align}
where the first term on the right hand side (RHS) of Eq.~\eqref{HS} is the von Neumann entropy $S^{\rm vN}:=-{\rm Tr}[\hat{\rho}(\theta)\ln\hat{\rho}(\theta)]$ of the system S.
Equation~\eqref{HS} contains the excess entropy  $S^{\mathrm{ex}}(\theta)$ defined as
\begin{align}\label{excess_entropy}
    S^{\mathrm{ex}}(\theta) :=-{\rm Tr}[\hat{\rho}\ln \hat{\rho}^{\rm ss} ] .
\end{align}
We should note that the housekeeping entropy $S^{\rm hk}$ to maintain the steady state is not discussed in the main text, because it can be decoupled from $S^{\mathrm{HS}}$.
See Appendix~\ref{housekeeping} for $S^{\rm hk}$.
Thus, the contribution of the Joule heating, which is $T S^{\rm hk}$, is safely distinguishable from the contribution discussed in the main text.

To discuss the performance of the engine, we introduce 
the one-cycle averaged entropy production defined as:
\begin{align}\label{A-def}
 \sigma :=  \frac{1}{2\pi}\int^{2\pi}_{0}   S^{\mathrm{HS}}(\theta) d\theta .
\end{align}
According to the inequality $S^{\mathrm{HS}}(\theta) \geq 0$, the entropy production is always positive semidefinite, i.e.\,$\sigma \geq 0$.
This $\sigma$ is related to the dissipative availability $A:=T\sigma$~\cite{Brandner-Saito,Hino2021}.
In other words, nonzero $\sigma$ means that the system is driven by a geometrical variable such as the BSN curvature.
Thus, $\sigma$ plays a key role in nonequilibrium thermodynamics.
Introducing the {\it mechanical work}
\begin{align}\label{W}
 W :=  \frac{1}{2\pi}\int^{2\pi}_{0} {\rm Tr} 
 \left[
  \hat{\rho}(\theta)\frac{ \partial \hat{H}(\lambda(\theta))}{\partial \lambda(\theta)}  \right] 
  \dot{\lambda}(\theta) d\theta 
\end{align}
with $\dot{\lambda}:=\frac{d}{d\theta}\lambda(\theta)$, 
we also have a scalar function
\begin{align}\label{A=U+V-W}
 Q: =W+T \sigma,
\end{align}
which satisfies the inequality
\begin{align}\label{thermo_2nd_law}
    W = Q - T\sigma \leq  Q ,
\end{align}
because $T\sigma\ge 0$.
The scalar quantity $Q$ is the avaliable energy corresponding to the absorbed heat of this engine.
Although $W$ can be negative in the definition of Eq.~\eqref{W}, it must be positive semidefinite if we restrict our interest to the case of $\dot{\lambda}\ge 0$ and ${\rm Tr}[\hat{\rho}(\theta)\partial \hat{H}(\lambda(\theta))/\partial \lambda]\ge 0$.
If these conditions are satisfied, $W$, $Q$ and Eq~\eqref{thermo_2nd_law} corerspond to the work done on the system through the modulation of the system Hamitonian $\hat{H}(\lambda(\theta))$, the absorbed heat and the second law of thermodynamics, respectively.   
However, the system cannot be perfectly periodic if the modulation satisfies the condition $\dot{\lambda}\ge 0$.
Indeed, the eigenvalue of $\hat{H}(\lambda(\theta))$ should increase after a one-cycle operation by an external agent, while a completely periodic operation must satisfy $\hat{H}(\lambda(\theta+2\pi))=\hat{H}(\lambda(\theta))$.\footnote{
If we are interested in the work done by the engine to the external environment, we need to focus on the situation ${\rm Tr}[\hat{\rho}\partial \hat{H}/\partial \lambda] \le 0$ in which the restored energy in $\hat{H}$ is released to perform the work to the environment. 
In this case the argument in the main text is almost unchanged except for the sign of $W$ as $W\le 0$.
Therefore, we only discuss the case of positive semidefinite $W$.
}
The possibility of the implementation of a perfectly periodic engine under the condition $\hat{H}(\lambda(\theta+2\pi))=\hat{H}(\lambda(\theta))$ will be discussed in Sec. \ref{sec:discussion}.
(See also Ref.~\cite{Yoshii22} for this case.)
Although we restrict our interest to the case of $W\ge 0$ in this paper,
it can be an engine to extract the work if we discuss the case of $W<0$, which is achieved for the perfectly periodic case~\cite{Yoshii22}.
Therefore, one can regard this paper as a proposal of a geometrical quantum engine to extract the work driven by the modulation of electrochemical potentials in the reservoirs and one parameter in Hamiltonian.


Let us introduce the ratio of the work $W$ to the available energy $Q$ as
\begin{align}\label{xi}
    \eta^{\rm eff} := \frac{W}{Q} = \frac{W}{W+T\sigma}.
\end{align}
From Eq.~\eqref{thermo_2nd_law}, $\eta^{\rm eff}$ satisfies $0 \leq \eta^{\rm eff} \leq 1$ if $\dot{\lambda}(\theta)\ge 0$ is satsified for any $\theta$. 
Thus, we call $\eta^{\rm eff}$ the effective efficiency because it is an indicator of the performance of the heat engine.
In the quasi-static limit ($\epsilon \to 0$), $\eta^{\rm eff}$ converges to $1$.
The scaled power defined as
\begin{align}\label{P}
    P := \epsilon W
\end{align}
converges to zero in this limit.
Note that $P$ does not have the dimension of power because we set the time scale by the dimensionless parameter $\epsilon$ under fixed $\Gamma$.
To obtain larger power, we need a higher speed of operation $\epsilon$, which leads to the effective efficiency $\eta^{\rm eff}$ becoming smaller.
Although $\eta^{\rm eff}$ differs from the conventional thermodynamic efficiency, $\eta^{\rm eff}$ is related to the entropy production during a cyclic operation of the engine, and can thus be regarded as a new efficiency for a thermodynamic engine as indicated by Refs.~\cite{Brandner-Saito,Hino2021}.

Before we proceed with the argument, we mention the possibility of the replacement of Eq.~\eqref{A-def} by
\begin{align}\label{A-def-repl}
\Delta S:&=S^{\rm HS}(\hat{\rho}(2\pi)||\hat{\rho}^{\rm ss}(2\pi))-S^{\rm HS}(\hat{\rho}(0)||\hat{\rho}^{\rm ss}(0)) \notag\\
&=\int_0^{2\pi}d\theta \dot{S}^{\rm HS}(\hat{\rho}(\theta)||\hat{\rho}^{\rm ss}(\theta)) .
\end{align}
This is the change of $S^{\rm HS}$ during a one-cycle operation. 
This $\Delta S$ is expected to be negative semidefinite if $S^{\rm HS}$ is the Kullback-Leibler (KL) divergence~\cite{Sagawa20,Petz86,Petz03,Ruskai02,Hiai11}.
We will discuss the proper choice of the entropy production for the dynamics in Sec.~\ref{sec:discussion}.

\subsection{Linear response regime}\label{sec:linear-response}

In this subsection, we consider the thermodynamics of the engine introduced in the previous subsection in the linear response regime for small $\epsilon$. 
See Appendix~\ref{app:slow-driving} for the general method used in this subsection.
When we assume that $\hat{\rho}^{\rm ss}$ is diagonalizable, there exists $(\hat{\rho}^{\rm ss})^{-1}$. 
As shown in Ref.~\cite{Yoshii22} the geometrical correction to $\hat{\rho}^{\rm ss}$ is negligible except for the initial relaxation process to reach a quasi-steady state.
In this case, we can expand
\begin{equation}\label{linear_expansion}
\hat{\rho}(\theta)=\hat{\rho}^{\rm ss}(\theta)+\epsilon \hat{\rho}^{(1)}(\theta) +O(\epsilon^2) .
\end{equation}
Substituting this into the expression for $\sigma$ in Eq.~\eqref{A-def}, with the help of Eqs.~\eqref{HS} and \eqref{excess_entropy}, we can expand $S^{\rm HS}$ as
\begin{align}
S^{\rm HS}(\hat{\rho}||\hat{\rho}^{\rm ss})
&=
\frac{1}{2}\epsilon^2{\rm Tr}
\left[ \hat{\rho}^{(1)}(\hat{\rho}^{\rm ss})^{-1}\hat{\rho}^{(1)} \right]
+O(\epsilon^3) ,
\end{align}
where we have used 
${\rm Tr}\hat{\rho}^{(1)}={\rm Tr}\hat{\rho}^{(2)}=0$.
Thus, we can rewrite $\sigma$ as
\begin{equation}\label{A-ad}
\sigma=\epsilon^2 \sigma^{(2)}+O(\epsilon^3),
\end{equation}
where $\sigma^{(2)}$ is expressed as
\begin{align}\label{A-ad-1}
\sigma^{(2)}&=\frac{1}{4\pi}\int_0^{2\pi} d\theta 
 {\rm Tr}[\hat{\rho}^{(1)}(\hat{\rho}^{\rm ss})^{-1}\hat{\rho}^{(1)}] .
\end{align}
As will be shown,
${\rm Tr}[\hat{\rho}^{(1)}(\hat{\rho}^{\rm ss})^{-1}\hat{\rho}^{(1)}]$ should be positive semidefinite.

The remaining problem is to evaluate $\hat{\rho}^{(1)}$ to obtain $\sigma^{(2)}$.
For this purpose we rewrite
 Eq.~\eqref{linear_expansion} as
\begin{align}\label{p-ad}
    |\hat{\rho} (\theta)\rangle \simeq |\hat{\rho}^{\mathrm{ss}}(\bm{\Lambda}(\theta))\rangle + \epsilon |\hat{\rho}^{(1)}(\bm{\Lambda}(\theta))\rangle +O(\epsilon^{2}) .
\end{align}
With the aid of Appendix \ref{app:slow-driving}, the second term on the RHS of Eq. (\ref{p-ad}) can be written as
\begin{align}\label{rho^1}
    |\hat{\rho}^{(1)}(\bm{\Lambda}(\theta))\rangle = 
    |\partial_\nu \hat{\rho}^{\rm ss}(\bm{\Lambda}(\theta)) \rangle \dot{\Lambda}_\nu(\theta) ,
 \end{align}   
 where we have introduced
 \begin{align}\label{partialnurho}
 |\partial_\nu\hat{\rho}^{\rm ss}(\bm{\Lambda}(\theta)\rangle
 :=\hat{K}^+(\bm{\Lambda}(\theta))
   \frac{\partial}{\partial \Lambda_\nu(\theta)}|\hat{\rho}^{\rm ss}(\bm{\Lambda}(\theta)) \rangle
   \end{align}
  using the pseudo-inverse $K^+(\bm{\Lambda}(\theta))$ \cite{inverse} of the transition matrix $\hat{K}(\bm{\Lambda}(\theta))$ and Eq.~\eqref{pn} for $n=1$. 
Here we define $\dot{\Lambda}_\mu=\frac{d}{d\theta}\Lambda_\mu$.
The definition of $\hat{K}^+(\bm{\Lambda}(\theta))$ is given as follows.
If  $\hat{K}(\bm{\Lambda})$ is diagonalizable, we can use the spectral decomposition as 
\begin{equation}\label{K-expand}
\hat{K}(\bm{\Lambda}) = \sum_{m} \varepsilon_{m}(\bm{\Lambda}) |r_{m}(\bm{\Lambda})\rangle \langle \ell_{m}(\bm{\Lambda})|,
\end{equation}
where $\varepsilon_{m}(\bm{\Lambda})$ is the eigenvalue.
$|r_{m}(\bm{\Lambda})\rangle$ and $\langle \ell_{m}(\bm{\Lambda})|$ are the corresponding right and left eigenvectors of $\hat{K}(\bm{\Lambda})$, respectively.
Then, $\hat{K}^{+}(\bm{\Lambda})$ is defined as
\begin{align}\label{K+}
    \hat{K}^{+}(\bm{\Lambda}) = \sum_{m\neq 0} \varepsilon_{m}(\bm{\Lambda})^{-1} |r_{m}(\bm{\Lambda})\rangle \langle \ell_{m}(\bm{\Lambda})| .
\end{align}

Thus, we can express $\hat{\rho}^{(1)}$ as
\begin{equation}\label{rho^{(1)}}
\hat{\rho}^{(1)}:=\partial_\nu \hat{\rho}^{\rm ss} \dot{\Lambda}_\nu ,
\end{equation}
where $\partial_\nu \hat{\rho}^{\rm ss}$ is the $n\times n$ matrix defined as
\begin{align}
\partial_\nu\hat{\rho}^{\rm ss}:=
\begin{pmatrix}
\partial_\nu\rho^{\rm ss}_{1} & \partial_\nu\rho^{\rm ss}_{2} & \cdots & \partial_\nu\rho^{\rm ss}_{n} \\
\partial_\nu\rho^{\rm ss}_{n+1} & \partial_\nu\rho^{\rm ss}_{n+2} & \cdots & \partial_\nu\rho^{\rm ss}_{2n} \\
\cdots &\cdots & \cdots & \cdots \\
\partial_\nu\rho^{\rm ss}_{n^2-n+1} & \partial_\nu\rho^{\rm ss}_{n^2-n+2} & \cdots & \partial_\nu\rho^{\rm ss}_{n^2} \\ 
\end{pmatrix}
\end{align} 
corresponding to $|\partial_\nu \hat{\rho}^{\rm ss}\rangle$ which has $n^2$ components.
Substituting Eq.~\eqref{rho^{(1)}} into Eq.~\eqref{A-ad-1} we obtain
\begin{equation}\label{Eq(30)}
\sigma^{(2)}=\frac{1}{2\pi}\int_0^{2\pi}d\theta
 g_{\mu\nu}(\bm{\Lambda}(\theta))
\dot{\bm{\Lambda}}_\mu(\theta)\dot{\bm{\Lambda}}_\nu(\theta),
\end{equation}
where the metric tensor $g_{\mu\nu}$ is given by
\begin{equation}\label{g_mn}
g_{\mu\nu}(\bm{\Lambda}):=
\frac{1}{2} {\rm Tr}
[\partial_\mu \hat{\rho}^{\rm ss}(\bm{\Lambda})(\hat{\rho}^{\rm ss}(\bm{\Lambda}))^{-1} \partial_\nu \hat{\rho}^{\rm ss}(\bm{\Lambda})] .
\end{equation}

There are two important properties of the metric tensor $g_{\mu\nu}$.
First, $g_{\mu\nu}$ is related to the Fisher information matrix~\cite{Fisher} and the Hessian matrix, since
Eq. \eqref{g_mn} can be rewritten as
\begin{align}\label{Fisher}
g_{\mu\nu}&= \frac{1}{2} {\rm Tr}[\hat{\rho}^{\rm ss} \partial_\mu \ln\hat{\rho}^{\rm ss} \partial_\nu \ln \hat{\rho}^{\rm ss} ] \notag\\
&=-\frac{1}{2}{\rm Tr}[\hat{\rho}^{\rm ss} \partial_\mu \partial_\nu \ln \hat{\rho}^{\rm ss}] , 
\end{align}
where we have used Eq.~\eqref{Fisher=Hessian}.
The expression on the first line of Eq.~\eqref{Fisher} is 
the Fisher information of $\hat{\rho}^{\rm ss}$ and the expression on the second line is the negative of the Hessian matrix of $\ln\hat{\rho}^{\rm ss}$.
Thus, the metric tensor can be expressed in terms of the Fisher information and Hessian matrix~\cite{Fisher}.
Second, $g_{\mu\nu}\dot{\Lambda}_\mu\dot{\Lambda}_\nu$ is positive semidefinite.
Indeed, it is straightforward to show that
\begin{equation}
\dot{\Lambda}_\mu g_{\mu\nu} \dot{\Lambda}_\nu
=\frac{1}{2}{\rm Tr}
\left[\hat{\rho}^{\rm ss} 
\left(
\dot{\Lambda}_\mu \hat{K}^+\frac{\partial}{\partial \Lambda_\mu}\ln \hat{\rho}^{\rm ss}
\right)^2 
\right]
\ge 0.
\end{equation}


Using the Cauchy-Schwartz inequality, 
the entropy production is bounded as
\begin{align}\label{CS}
   \sigma^{(2)} \geq   \mathcal{L}^{2},
\end{align}
where
\begin{align}\label{L}
    \mathcal{L} :&= \oint_{\partial\Omega} \sqrt{ g_{\mu\nu}(\bm{\Lambda}) d\Lambda_{\mu} d\Lambda_{\nu}} \notag\\
&:= \frac{1}{2\pi} \int_0^{2\pi} d\theta \sqrt{ g_{\mu\nu}(\bm{\Lambda}) \dot{\Lambda}_{\mu}(\theta)\dot{\Lambda}_{\nu}(\theta)}
\end{align}
is the thermodynamic length corresponding to the length along the closed trajectory $\partial\Omega$ surrounding the domain $\Omega$ of the Riemannian manifold with the metric $g_{\mu\nu}(\bm{\Lambda})$.
The inequality \eqref{CS} has an identical form to that of the system coupled only to one reservoir~\cite{Brandner-Saito}.
The equality in Eq.~\eqref{CS} holds when $g_{\mu\nu}(\bm{\Lambda}(\theta)) \dot{\Lambda}_{\mu}(\theta) \dot{\Lambda}_{\nu}(\theta)$ is a non-negative $\theta$-independent constant. 
This equality cannot be achieved if the BSN curvature is meaningful, because $g_{\mu\nu}(\bm{\Lambda}(\theta)) \dot{\Lambda}_{\mu}(\theta) \dot{\Lambda}_{\nu}(\theta)$ should be a $\theta$-dependent variable when the trajectory in the parameter space makes a closed loop to generate BSN curvature.

Let us consider the leading-order contribution to to the work $W$, written as $W=\mathcal{W}+O(\epsilon)$, where
\begin{align}\label{W-ad}
    \mathcal{W} := \oint_{\partial\Omega} \mathcal{A}_{\mu}(\bm{\Lambda}) d\Lambda_{\mu}
\end{align}
is the adiabatic work expressed as the line integral of the thermodynamic vector potential 
\begin{align}\label{vector-potential}
    \mathcal{A}_{\mu}(\bm{\Lambda}) 
    :&=\delta_{\mu,0} {\rm Tr}\left[ \frac{\partial\hat{H}}{\partial \lambda}  \hat{\rho}^{\mathrm{ss}}(\bm{\Lambda})\right]
\end{align}
along the trajectory $\partial \Omega$ of parameter control \cite{Brandner-Saito}.
Note that $\mathcal{A}_{\mu}(\bm{\Lambda})$ corresponds to the BSN vector in adiabatic pumping processes~\cite{sinitsyn1,sinitsyn2}.
We also note that the adiabatic work $\mathcal{W}$ is independent of $\epsilon$.
We stress that the adiabatic work in Eq~\eqref{W-ad} is expressed as the geometrical vector potential in Eq.~\eqref{vector-potential} as was alluded to below Eq.~\eqref{W}.
Therefore, if $\mathcal{W}$ in Eq.~\eqref{W-ad} is finite, we extract the work from the quantum geometrical engine under the control of external electrochemical potentials $\mu^{\rm L}$ and $\mu^{\rm R}$.

The adiabatic work is related to the flux penetrating the area $\Omega$ surrounded by $\partial\Omega$.
By using the Stokes theorem, one can rewrite the adiabatic work as 
\begin{equation}\label{W=oint}
\mathcal{W}=\oint_{\partial \Omega}  \mathcal{A} 
=
\int_{\Omega} d\mathcal{A} .
\end{equation}
Here we define the 1-form $\mathcal{A}=\mathcal{A}_\mu d\Lambda_\mu$. 
Introducing the thermodynamic curvature $F_{\mu\nu}$  
\begin{equation}
F_{\mu\nu}:=\frac{\partial}{\partial \Lambda_\mu} A_\nu-\frac{\partial}{\partial \Lambda_\nu}   A_\mu, 
\label{Gdef}
\end{equation}
we can rewrite the 2-form $d\mathcal{A}$ as
\begin{equation}\label{2-form_dA}
d\mathcal{A}=\frac{1}{2}F_{\mu\nu}d\Lambda^\mu \wedge d\Lambda^\nu .
\end{equation}
This $d\mathcal{A}$ is directly related to the thermodynamic axial field $\vec{\mathscr{B}}=(\mathscr{B}_\mu)$ and the thermodynamic flux $\Phi_{\rm TD}$ as
\begin{align}\label{B=rotA}
\mathscr{B}_\mu&:=\epsilon_{\mu\nu\rho} F_{\nu\rho} , \\
\Phi_{\rm TD}:&=\int_\Omega \vec{\mathscr{B}}\cdot d\vec{S}
 ,
 \label{flux}
\end{align}
where $dS_\mu:=\frac{1}{2}\epsilon_{\mu\nu\rho}d\Lambda_\nu \wedge d\Lambda_\rho$ with 
 the Levi-Civita symbol $\epsilon_{\mu\nu\rho}$, i.e. $\epsilon_{\mu\nu\rho}=1$ for an even permutation of $(\mu,\nu,\rho)$, $\epsilon_{\mu\nu\rho}=-1$ for an odd permutation of $(\mu,\nu,\rho)$ and $\epsilon_{\mu\nu\rho}=0$ otherwise. 
From Eqs.~\eqref{vector-potential}, \eqref{Gdef} and \eqref{B=rotA}, $\mathscr{B}_0=0$, and thus, $\vec{\mathscr{B}}$ can be regarded as a two-dimensional vector which has only the components $\mathscr{B}_1$ and $\mathscr{B}_2$. 

From the definitions Eqs.~(\ref{Gdef}) and \eqref{B=rotA}, $\mathscr{B}_\mu$ satisfies Gauss' law as 
\begin{align}\label{divergence_free}
\partial_\mu {\mathscr{B}}_\mu=0.
\end{align}
With the aid of Eqs.~\eqref{W=oint}, \eqref{Gdef}, \eqref{2-form_dA} and \eqref{B=rotA}, Eq.~\eqref{flux} can be rewritten as
\begin{equation}
\Phi_{\rm TD}=\oint_{\partial \Omega} \mathcal{A}_\mu d\Lambda_\mu=\mathcal{W} .
\end{equation}
Thus, the thermodynamic flux $\Phi_{\rm TD}$ reduces to the line integral along the closed path $\partial \Omega$, which is the adiabatic work $\mathcal{W}$.

The average power can be expressed as
\begin{align}\label{P-ad}
    P = \epsilon \mathcal{W} + O(\epsilon^2)
\end{align}
for small $\epsilon$.

By using the equality (\ref{A-ad}), the effective efficiency $\eta^{\rm eff}$ introduced in Eq.~\eqref{xi} is rewritten as
\begin{align}\label{xi_eq}
    \eta^{\rm eff} &= 
\frac{\mathcal{W}}{\mathcal{W}+\epsilon^2 T \sigma^{(2)}}
\notag\\
&=1 - \epsilon^2 T \frac{\sigma^{(2)}}{\mathcal{W}} + O(\epsilon^{2}) . 
\end{align}
Using Eq.~\eqref{CS}, the relation \eqref{xi_eq} can be rewritten as
\begin{equation}\label{xi_inequality}
1-\eta^{\rm eff} \ge \epsilon^2 \frac{T\mathcal{L}^2}{\mathcal{W}}=\epsilon^3 \frac{T\mathcal{L}^2}{P} ,
\end{equation}
where we have used Eq.~\eqref{P} for the final expression.
This relation tells us that the decrement of the effective efficiency is bounded by the thermodynamic length $\mathcal{L}$ and $\mathcal{W}$ or the power $P$.

It is straightforward to discuss the trade-off relation between the power $P$ and $\eta^{\rm eff}$ as follows.
Substituting Eq.~\eqref{P-ad} into Eq.~\eqref{xi_inequality} we can rewrite it as
$\eta^{\rm eff}\le 1-P^2T\mathcal{L}^2/\mathcal{W}^3$.
Rewriting this as an expression for $P$ we obtain
\begin{equation}
P\le \frac{W^{3/2}}{\mathcal{L}}\displaystyle\sqrt{\frac{1-\eta^{\rm eff}}{T}} ,
\end{equation}
which is equivalent to the trade-off relation in Refs.~\cite{Brandner-Saito,Hino2021}.

It is obvious that $\eta^{\rm eff}$ in Eq.~\eqref{xi_eq} becomes 1
in the adiabatic limit ($\epsilon\to 0$).
Note that the conventional thermodynamic efficiency does not exist in our system because there is no heat current between two reservoirs.
Therefore, the bound of the efficiency of this engine is not the Carnot efficiency,
even if $\eta^{\rm eff}$ reaches the maximum value.
Nevertheless, we do not consider any cost of controlling the system Hamiltonian, and electrochemical potentials, which is most important for the realization of this engine.

\section{APPLICATION to a quantum dot system}\label{sec:appli}

\begin{figure}
\centering
\includegraphics[clip,width=8cm]{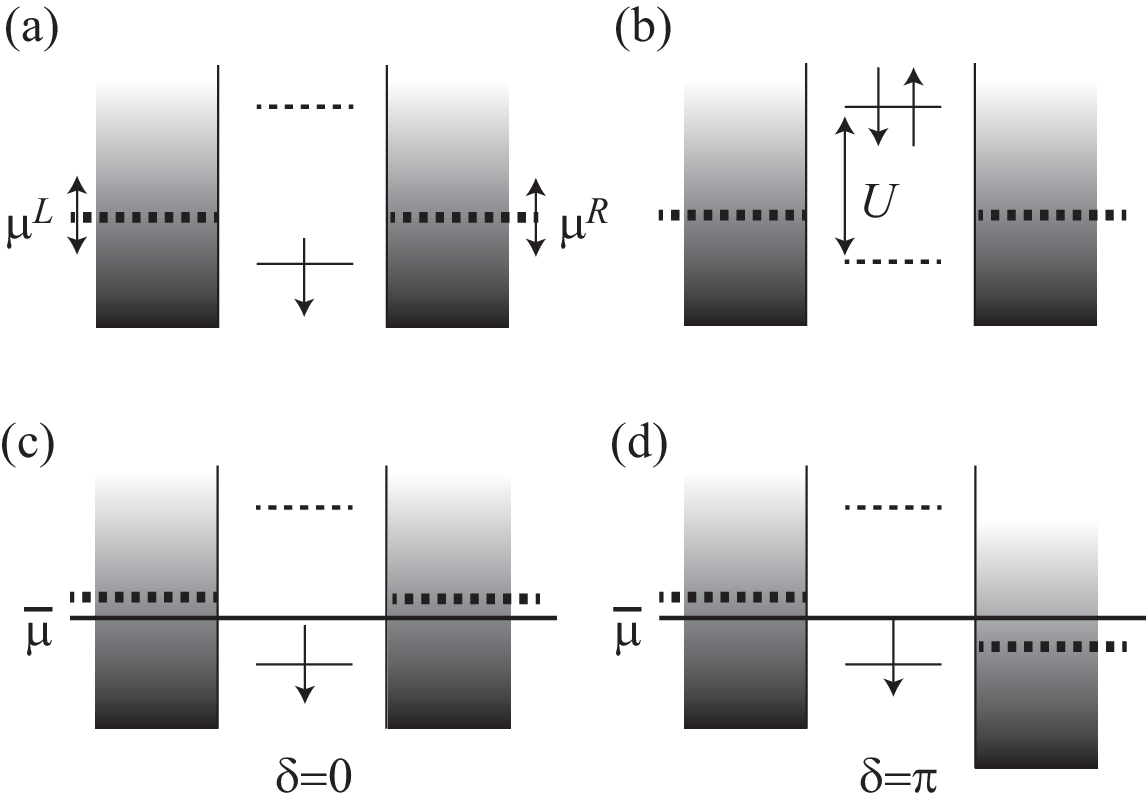}
\caption{
(a) 
Schematics of the modulation of the Anderson model in which two reservoirs are connected to a quantum dot.
We modulate the electrochemical potentials in the left ($\mu^{\rm L}(\theta)$) and right  ($\mu^{\rm R}(\theta)$) reservoirs. 
(b)
We also modulate the Coulomb repulsion $U$ inside the quantum dot as a tunable parameter.
There is another contol parameter $\delta$, the phase difference of the modulation of the two electrochemical potentials.
Figure (c) is a schematic for $\delta=0$ and Fig. (d) is that for $\delta=\pi$.    
}
\label{figAM}
\end{figure}

In this section, we apply the general framework developed in the previous section to the Anderson model for a quantum dot (QD)
 in which a single dot is coupled to two electron reservoirs (see Fig. \ref{figAM} for a set of schematics of time evolution of our system).
 

\subsection{Anderson model for a quantum dot}

The total system consists of the single-dot system and baths (reservoirs).
Thus, the total Hamiltonian $\hat{H}^{\rm tot}$ is written as 
\begin{align}\label{H_tot}
\hat{H}^{\rm tot}:&=\hat{H}+\hat{H}^{\rm r}+\hat{H}^{\rm int} ,
\end{align}
where the system Hamiltonian $\hat{H}$, reservoir Hamitonian $\hat{H}^{\rm r}$ and interaction Hamiltonian $\hat{H}^{\rm int}$ are, respectively, given by
\begin{align}\label{H_s}
\hat{H}&=\sum_\sigma \epsilon_0 \hat{d}_\sigma^\dagger \hat{d}_\sigma+U(\theta) \hat{n}_\uparrow \hat{n}_\downarrow ,
\\
\label{H_bath}
\hat{H}^{\rm r}&=\sum_{\alpha,k,\sigma}\epsilon_k \hat{a}_{\alpha,k,\sigma}^\dagger \hat{a}_{\alpha,k,\sigma} ,
\\
\hat{H}^{\rm int}&=\sum_{\alpha,k,\sigma}V_\alpha \hat{d}_\sigma^\dagger \hat{a}_{\alpha,k,\sigma}+{\rm h.c.},
\end{align}
where $\hat{a}_{\alpha, k,\sigma}^\dagger$ and $\hat{a}_{\alpha, k,\sigma}$ are, respectively, the creation and annihilation operators  for the electron in the reservoirs $\alpha$$=$(L or R) with wave number $k$, energy $\epsilon_k$, and spin $\sigma=(\uparrow$ or $\downarrow$).
Moreover,  $\hat{d}^\dagger_\sigma$ and $\hat{d}_\sigma$ are those in the QD, and  
$\hat{n}_{\sigma}=\hat{d}^\dagger_{\sigma}\hat{d}_{\sigma}$. 
$U(\theta)$ and $V_\alpha$ are, respectively, the time-dependent electron-electron interaction in the QD and the transfer energy between the QD and the reservoir $\alpha$. 
We adopt a model in the wide-band limit for the reservoirs.
In this paper, the line width is given by $\Gamma=\pi \mathfrak{n} V^2$, where $V^2=V_L^2+V_R^2$ and $\mathfrak{n}$ is the density of states in the reservoirs.

In this paper, we consider geometrical pumping caused by an adiabatic modulation of the parameters. 
As stated, we adiabatically control $\mu^{\mathrm{L}}$ and $\mu^{\mathrm{R}}$ with the condition 
$\overline{\mu}:=\overline{\mu^{\mathrm{\alpha}}}$ and $U(\theta)$, and fix the other parameters.
The modulation of $U(\theta)$ is described as
\begin{equation}\label{U(theta)}
U(\theta)=U_0\lambda(\theta), \quad \lambda(\theta)=\theta+r_\lambda \cos\theta ,
\end{equation}
where we have assumed $|r_\lambda|\le 1$.

The Anderson model for a quantum dot should have the following four states (corresponding to $n=4$ in the previous section): doubly occupied, singly occupied with an up-spin, singly occupied with a down-spin and empty.
Therefore, the density matrix should be expressed as a $4\times 4$ matrix.
As is shown in Ref.~\cite{Yoshii13}, however, the density matrix of the quantum master equation of the Anderson model within the wide-band approximation is reduced to a four component matrix
\begin{equation}\label{density_matrix}
\hat{\rho}=
\begin{pmatrix}
\rho_d & 0 & 0& 0 \\
0 & \rho_\uparrow & 0 & 0 \\
0 & 0 & \rho_\downarrow& 0 \\
0 & 0 & 0 & \rho_e \\
\end{pmatrix}
 ,
\end{equation}
where $\rho_d,\rho_\uparrow,\rho_\downarrow$, and $\rho_e$ correspond to probabilities of the doubly occupied state, singly occupied state with up-spin, singly occupied state with down-spin, and empty state, respectively. 
This means that the model is not fully quantum-mechanical but quasi-classical.

Since Eq.~\eqref{density_matrix} is diagonal, $|\hat{\rho}\rangle$ also has only four components and the transition matrix $\hat{K}(\bm{\Lambda}(\theta))$ in Eq.~\eqref{master} in the wide band approximation is given by the $4\times 4$ matrix (see Appendix~\ref{K-matrix} and Ref.~\cite{Nakajima15} for the derivation)
\begin{equation}\label{Ville_9_0806}
\hat{K}(\bm{\Lambda}(\theta))
=\begin{pmatrix}
-2f_-^{(1)} & f_+^{(1)} & f_+^{(1)} & 0 \\
f_-^{(1)}& -f_-^{(0)}-f_+^{(1)} & 0 & f_+^{(0)} \\
f_-^{(1)}& 0 & -f_-^{(0)}-f_+^{(1)}  & f_+^{(0)} \\
0 & f_-^{(0)} & f_-^{(0)} & -2f_+^{(0)} \\
\end{pmatrix} ,
\end{equation}
where 
we have introduced
\begin{align}
\label{f_+}
f_+^{(j)}:&=f_L^{(j)}(\mu^L,U)+f_R^{(j)}(\mu^R,U) \\
f_-^{(j)}:&=2-\{f_L^{(j)}(\mu^L,U)+f_R^{(j)}(\mu^R,U) \}
\label{f_-}
\end{align}
with the Fermi distribution
\begin{equation}\label{Fermi}
f_\alpha^{(j)}(\mu^\alpha(\theta),U(\theta)):=\frac{1}{1+e^{\beta (\epsilon_0+j U(\theta)-\mu^\alpha(\theta))}}
\end{equation}
in the reservoir $\alpha(=L~{\rm or}~ R)$.
Note that Eqs. \eqref{f_+} and \eqref{f_-} satisfy the relation 
\begin{equation}\label{sum_rule}
f_+^{(j)}+f_-^{(j)}=2
\end{equation}
for $j=0$ and 1.
Here, the pseudo-inverse $\hat{K}^+$ for the Anderson model is explicitly written in Appendix \ref{sec:K^+}.

It is straightfoward to obtain the eigenvalues of $\hat{K}(\bm{\Lambda}(\theta))$ in Eq.~\eqref{Ville_9_0806} as
\begin{align}
\varepsilon_0&=0, \label{0_eigen}\\
\varepsilon_1&=-(f_+^{(0)}+f_-^{(1)}), \label{1_eigen}\\
\varepsilon_2&=-(f_-^{(0)}+f_+^{(1)}), \label{2_eigen}
\\ 
\varepsilon_3&=-4.
\label{3_eigen}
\end{align}

The left and right eigenfunctions corresponding to $\varepsilon_0=0$ in Eq.~\eqref{0_eigen} are given by
\begin{equation}\label{left_zero}
\langle \ell_0|=(1, 1, 1, 1) ,
\end{equation}
and 
\begin{equation}\label{right_zero}
|r_0\rangle 
=
\frac{1}{2(f_+^{(0)}+f_-^{(1)})}
\begin{pmatrix}
f_+^{(0)}f_+^{(1)} \\[0.5em]  
f_+^{(0)} f_-^{(1)}\\[0.5em]   
f_+^{(0)} f_-^{(1)}\\[0.5em]  
f_-^{(0)} f_-^{(1)}
\end{pmatrix},
\end{equation}
respectively.
Because of Eq.~\eqref{steady_condition} there is the trivial relation $|r_0\rangle =|\hat{\rho}^{\rm ss}\rangle$ for the diagonal element of the density matrix.
Note that $|r_0\rangle$ satisfies $\langle\ell_0|r_0\rangle={\rm Tr}\hat{\rho}^{\rm ss}=1$ for Eq.~\eqref{density_matrix}.

The left and right eigenfunctions corresponding to $\varepsilon_1$ in Eq.~\eqref{1_eigen} are given by
\begin{equation}
\label{left_1}
\langle \ell_1 |
=
2
\begin{pmatrix}
f_-^{(1)}, &
\frac{-f_+^{(0)}+f_-^{(1)}}{2}, &
\frac{-f_+^{(0)}+f_-^{(1)}}{2},&
-f_+^{(0)}
\end{pmatrix},
\end{equation}
and
\begin{equation}\label{right_1}
|r_1\rangle
=
\frac{1}{(f_+^{(0)}+f_-^{(1)})(f_-^{(0)}+f_+^{(1)})}
\begin{pmatrix}
f_+^{(1)}\\[0.5em]
\frac{-f_+^{(0)}+f_-^{(1)}}{2} \\[0.5em]
\frac{-f_+^{(0)}+f_-^{(1)}}{2}  \\[0.5em]
-f_-^{(0)}
\end{pmatrix}.
\end{equation}

The left and right eigenfunctions corersponding to $\varepsilon_2$ in Eq.~\eqref{2_eigen} are
\begin{equation}\label{left_2}
\langle \ell_2 |=
2(0, 1, -1, 0),
\end{equation}
and
\begin{equation}\label{right_2}
|r_2\rangle
=\frac{1}{4}
\begin{pmatrix}
0\\
1\\
-1\\
0
\end{pmatrix}
,
\end{equation}
respectively.

The left and right eigenfunctions corresponing to $\varepsilon_2$ in Eq.~\eqref{3_eigen} are 
\begin{equation}\label{left_3}
\langle \ell_3|
  =
\begin{pmatrix}
 f_-^{(0)}f_-^{(1)},&
-f_-^{(0)}f_+^{(1)},&
-f_-^{(0)}f_+^{(1)},&
 f_+^{(0)}f_+^{(1)}
\end{pmatrix},
\end{equation}
and
\begin{equation}\label{right_3}
|r_3\rangle
=
 \frac{1}{2(f_-^{(0)}+f_+^{(1)})}
\begin{pmatrix}
1 \\[0.5em]
-1\\[0.5em]
-1\\[0.5em]
1
\end{pmatrix}.
\end{equation}


\subsection{Explicit calculation}


In this subsection, we calculate the thermodynamic quantities discussed in Sec. \ref{sec:general} numerically with the aid of the detailed properties of the Anderson model explained in Appendix~\ref{app:detailed_Anderson} . 
Although Ref.~\cite{Abiuso20} discussed the optimal path for high efficiency and work, we simply adopt the following control of the set of parameters 
$\bm{\Lambda}(\theta)$ as
\begin{align}\label{3rd_protocol}
    \lambda(\theta) &= \theta+ r_{\lambda} \cos\theta, \\
    \frac{\mu^{\mathrm{L}}(\theta)}{\overline{\mu}} &= 1 + r_\mu \sin \theta, \\
    \frac{\mu^{\mathrm{R}}(\theta)}{\overline{\mu}} &=1+ r_T \sin[\theta+ \delta], 
\end{align}
where $\delta$  is the phase difference between the electrochemical potentials in the left and right reservoirs.

If we take $\delta \neq 2n \pi$ with an integer $n$, the electrochemical potential difference between two reservoirs remains finite. 
For the explicit calculation, we set 
\begin{equation}
 r:= r_\lambda=r_\mu=r_T
\end{equation}
for simplicity.
For the explicit calculation, we fix the parameters  $\beta\epsilon _0 =\beta \overline{\mu}= \beta U_0 = 0.1$. 
By recasting $|\rho^{\rm{ss}}(\bm{\Lambda})\rangle=|r_0\rangle$ (Eq.\,(\ref{right_zero})) into the matrix form 
and plugging it into Eqs.\,(\ref{Eq(30)}), (\ref{g_mn}) and (\ref{L}) we can calculate $\sigma^{(2)}$ and $\mathcal{L}$. 
We note that it is almost independent of the choice of the initial condition. 
The adiabatic work (\ref{W-ad}) can also be calculated easily by substituting Eq.\.(\ref{F3}) into Eq.\,(\ref{vector-potential}) for the Anderson model. 
As can be seen from the explicit form, the adiabatic work is not sensitive to the choice of the initial condition. 


For the Anderson model, all the matrices involved are diagonalizable $4\times 4$ matrices.
Therefore, we can calculate the one-cycle average entropy production $\sigma^{(2)}$ and the square of the thermodynamic length $\mathcal{L}^2$ introduced in Eq.~\eqref{L} for the protocol in Eq.~\eqref{3rd_protocol}. 
To obtain the concise results and check the validity of numerical results we also develop a perturbative treatment under the condition $\beta U_0\ll 1$.
See Appendix~\ref{app:perturbed_Anderson} for details of this calculation.

\begin{figure}
\centering
\includegraphics[clip,width=8cm]{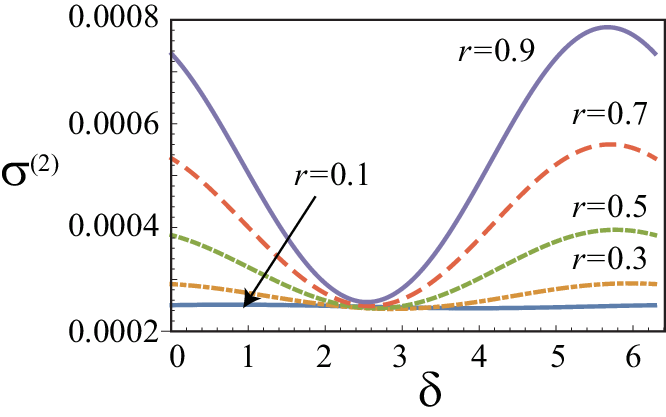}
\caption{
Plots of one-cycle averaged entropy production $\sigma^{(2)}$ against the phase difference $\delta$ for various $r$. 
We set $\epsilon=0.1$, $\beta U_0=0.1$, $\beta \epsilon_0=0.1$, and $\beta \bar \mu =0.1$.
}
\label{fig:oce}
\end{figure}

\begin{figure}
\centering
\includegraphics[clip,width=8cm]{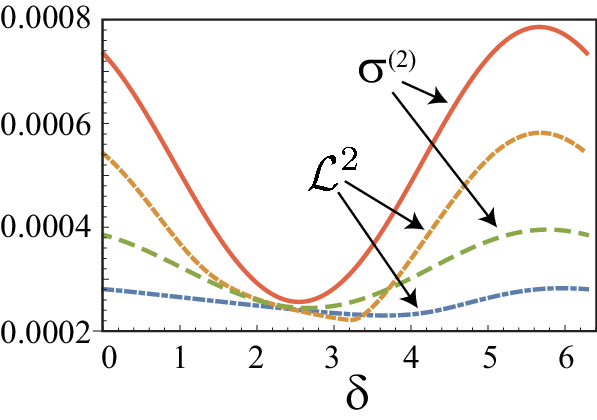}
\caption{
Plots of $\mathcal{L}^2$ (brown dashed lines for $r=0.9$ and dot-dashed line for $r=0.5$) and one-cycle averaged entropy production $\sigma^{(2)}$ (solid line for $r=0.9$ and green dashed line for $r=0.5$) against the phase difference $\delta$. 
We set $\epsilon=0.1$, $\beta U_0=0.1$, $\beta \epsilon_0=0.1$, and $\beta \bar \mu =0.1$.
}
\label{fig:th_l}
\end{figure}

We plot $\sigma^{(2)}$ against $\delta$ for various $r$ in Fig.~\ref{fig:oce}.
As we can see, $\sigma^{(2)}$ increases as $r$ increases.
We plot $\mathcal{L}^2$  and $\sigma^{(2)}$ as functions of the phase difference $\delta$ for $r=0.5$ and $r=0.9$ (see Fig.~\ref{fig:th_l}).
Figure \ref{fig:th_l} is consistent with the inequality Eq.~\eqref{CS}.
 


\begin{figure}
\centering
\includegraphics[clip,width=8cm]{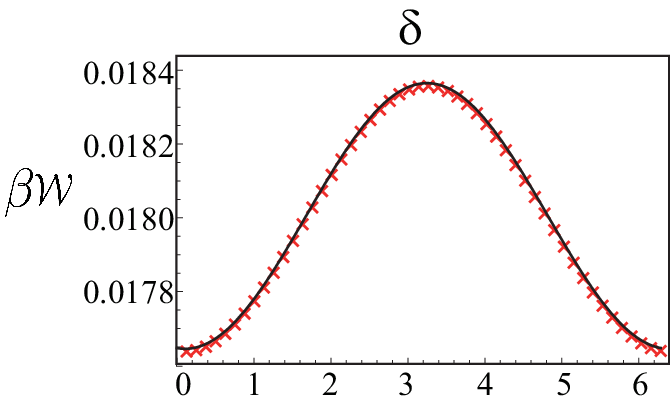}
\caption{Plot of the dimensionless adiabatic work $\beta \mathcal{W}$ as a function of the phase difference $\delta$ for $r=0.9$ (crosses).
The blue solid line expresses a fitting by a sinusoidal function. 
We set $\epsilon=0.1$, $\beta U_0=0.1$, $\beta \epsilon_0=0.1$, and $\beta \bar \mu =0.1$.
 }
\label{fig:work}
\end{figure}


\begin{figure}
\centering
\includegraphics[clip,width=8cm]{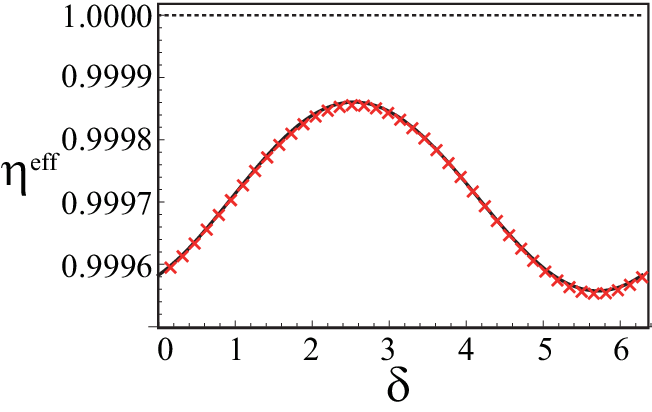}
\caption{Plot of the effective efficiency $\eta ^{\mathrm{eff}}$ as a function of the phase difference $\delta$ for $r=0.9$ with fixed $\epsilon=0.1$ (solid line), where the black and crosses are the upper bound $\eta^{\rm eff}=1$ and fitting by a sinusoidal function, respectively. 
We set $\beta U_0=0.1$, $\beta \epsilon_0=0.1$, and $\beta \bar \mu =0.1$.
}
\label{fig:efficiency}
\end{figure}

\begin{widetext}


The adiabatic work $\mathcal{W}$ defined in Eq.~\eqref{W-ad} can also be calculated as in Appendices \ref{app:detailed_Anderson} and \ref{app:perturbed_Anderson}.
The plot of $\beta\mathcal{W}$ versus $\delta$ for various $r$ is presented in Fig.~\ref{fig:work}.
It is noteworthy that the adiabatic work reaches a maximum at $\delta=\pi$, where the geometrical contribution from the BSN curvature does not exist.
On the other hand, $\mathcal{W}$ attains the minimum at $\delta=2n\pi$ with an integer $n$.
The functional form $\beta\mathcal{W}$ can be fitted by a sinusoidal function of $\delta$.

The effective efficiency $\eta^{\rm eff}$ defined in Eq.~\eqref{xi_eq} as a function of $\delta$ for fixed $\epsilon=0.1$ is presented in Fig.~\ref{fig:efficiency}.
Although $\eta^{\rm eff}$ is close to unity for all $\delta$, 
the maximum value of $\eta^{\rm eff}$ at $\delta=\pi$ remains below unity for finite $\epsilon$.
The functional form of $1-\eta^{\rm eff}$ can be also fitted to a sinusoidal function.

We also plot the two-dimesnsional thermodynamic axial field $\vec{\mathscr{B}}$ in the plane of $\mu^{\rm L}/\overline{\mu}$ and $\mu^{\rm R}/\overline{\mu}$ (see Fig,~\ref{fig:B_field}).
For both cases, $\vec{\mathscr{B}}$ flows in the direction of increasing $\mu^{\rm L}$ and decreasing $\mu^{\rm R}$.


\begin{figure}
\centering
\includegraphics[clip,width=16cm]{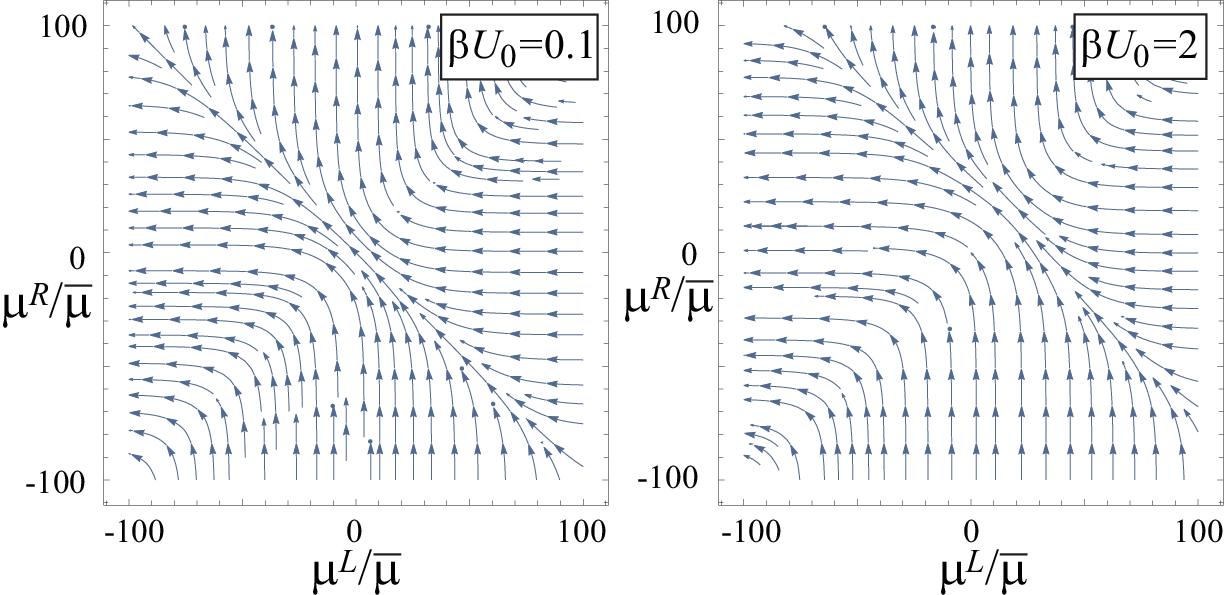}
\caption{Plots of the thermodynamic axial fields $\vec{\mathscr{B}}$ in the plane of $\mu^{\rm L}/\overline{\mu}$ and $\mu^{\rm R}/\overline{\mu}$ for $\beta U_0=0.1$ (left) and $\beta U_0=2$ (right). }
\label{fig:B_field}
\end{figure}

\end{widetext}

\section{Discussion}\label{sec:discussion}

So far, we have assumed the conditions $\dot{\lambda}\ge 0$ and ${\rm Tr}[\hat{\rho}(\theta)\partial \hat{H}(\lambda(\theta))/\partial \lambda]\ge 0$ to ensure the positive semidefinite property of $W$.
This means that the system Hamiltonian is not perfectly periodic, i.e. $\hat{H}(\lambda(\theta+2\pi))\ne \hat{H}(\lambda(\theta))$ in our analysis.
In this section, we mainly discuss what happens if we remove these requirements. 

\begin{figure}
\centering
\includegraphics[clip,width=8cm]{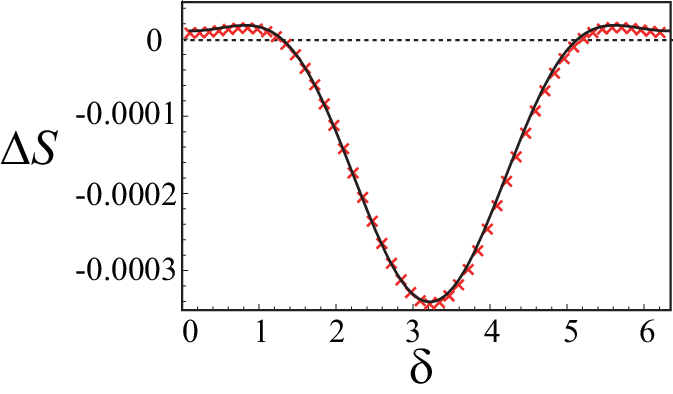}
\caption{Plot of $\Delta S$ in Eq.~\eqref{Delta_S>0} against $\delta$ (blue solid line). The red dashed lines is the fitting by a sinusoidal function with the second harmonic.
We set $\epsilon=0.1$, $r=0.9$, $\beta U_0=0.1$, $\beta \epsilon_0=0.1$, and $\beta \bar \mu =0.1$.
}
\label{fig:Delta_S}
\end{figure}

\begin{figure}
\centering
\includegraphics[clip,width=8.5cm]{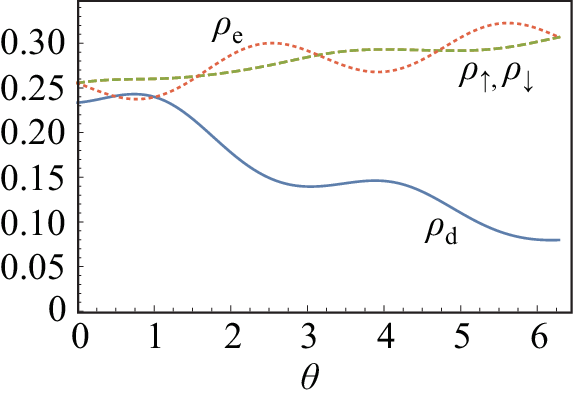}
\caption{Time evolution of the elements $\rho_d$, $\rho_\uparrow$, $\rho_\downarrow$ and $\rho_e$ of the density matrix in the first cycle.
We set $\epsilon=0.1$, $r=0.9$, $\beta U_0=0.1$, $\beta \epsilon_0=0.1$, and $\beta \bar \mu =0.1$.
}
\label{fig:density_matrix}
\end{figure}

First, we discuss whether $\Delta S$ introduced in Eq.~\eqref{A-def-repl} can be used. 
Indeed, $\Delta S$ is directly connected to the entropy production in a one-cycle operation.
If the conditions $\dot{\lambda}\ge 0$ and ${\rm Tr}[\hat{\rho}\partial \hat{H}/\partial \lambda]\ge 0$ are satisfied, $\Delta S$ given by
\begin{equation}\label{Delta_S>0}
\Delta S=
\epsilon^2\left[
g_{\mu\nu}(2\pi)\dot{\Lambda}_\mu(2\pi)\dot{\Lambda}_\nu(2\pi)- g_{\mu\nu}(0)\dot{\Lambda}_\mu(0)\dot{\Lambda}_\nu(0)
\right]
\end{equation}
in the linear response regime is expected to be negative semidefinite, because $\dot{S}^{\rm HS}\le 0$ is expected to be satisfied under any CPTP process~\cite{Sagawa20,Petz86,Petz03,Ruskai02,Hiai11}.
It is, however, surprising that $\Delta S$ is positive for small $\delta$ and for $\delta$ near $2\pi$ 
(see Fig.~\ref{fig:Delta_S}).
Since the dynamics satisfies CPTP properties (see Fig.~\ref{fig:density_matrix}~\footnote{
Figure~\ref{fig:density_matrix} plots the time evolution of the elements of $\hat{\rho}(\theta)$ as $\rho_{e}$, $\rho_{\downarrow}$, $\rho_{\uparrow}$ and $\rho_{d}$.
This figure clearly supports the positivity of all elements, therefore preserving the complete positivity argument.
}), this means that $S^{\rm HS}$ is not the KL divergence.
Namely the monotonicity of the KL divergence is only valid for the relaxation dynamics with fixed $\theta$, while $S^{\rm HS}$ evolves without the relaxation to $\hat{\rho}^{\rm SS}(\theta)$ during the dynamics.

Let us explore what happens if we consider the perfectly periodic case $\hat{H}(\lambda(\theta))=\hat{H}(\lambda(\theta+2\pi))$ in which $\dot{\lambda}(\theta)$ and $W$ are not positive semidefinite.
In this case, the effective efficiency $\eta^{\rm eff}$ and power $P$ cannot be well defined.
Nevertheless, most of our arguments, except for the ones concerning $\eta^{\rm eff}$ and $P$, in this paper can be used even for this perfectly periodic case.
See Ref.~\cite{Yoshii22}.

Although we cannot ensure $W\ge 0$ for an arbitrary setup with perfectly periodic modulation, we can choose parameters satisfying $W\ge 0$.
If we only consider such a situation, we can safely analyze $\eta^{\rm eff}$ and $P$, and $\eta^{\rm eff}$ can easily reach unity.
While this might be an unrealistic situation, it is an interesting one to consider.    

Let us consider the case where we control the temperature of one of the reservoirs.
If $\dot{S}^{\rm HS}$ is negative semidefinite, as in Refs.~\cite{Brandner-Saito,Hino2021}, we can use the dissipative availability $A$ defined by 
\begin{equation}
A:=\int_0^{2\pi}d\theta \Theta(\theta)\dot{S}^{\rm HS}(\theta),
\end{equation}
 where $\Theta(\theta)$ is the modulated temperature of one of the reservoirs satisfying $\overline{\Theta}=T$, as the basis of thermodynamic engine.
 However, as is shown in Fig.~\ref{fig:Delta_S} with Eq.~\eqref{Delta_S>0}, we cannot ensure the negative semidefinite property of $A$. 
Therefore, the problem of modulating the temperature is non-trivial. 
Analyzing this setup is one of our future tasks.


Finally, let us discuss the Cram\'{e}r-Rao bound~\cite{Cramer46,Rao45,Ito2020} which is associated with Fisher information. 
In fact, it is straightforward to derive the Cram\'{e}r-Rao bound as ${\rm Tr}[\delta\Lambda_\mu\delta \Lambda_\nu \hat{\rho}^{\rm ss}(\theta)]=\delta\Lambda_\mu\delta \Lambda_\nu\ge g_{\mu\nu}^{-1}$ with $\delta \Lambda_\mu:=\Lambda_\mu(\theta)-\overline{\Lambda_\mu(\theta)}$, but its physical meaning is unclear. 
Clarifying the meaning of the bound will be a topic of our future research.


\section{Concluding Remarks}\label{sec:conclusion}

In this paper, we successfully extended the geometrical thermodynamics formulated in Refs. \cite{Brandner-Saito,engine,Hino2021} to a quantum system coupled to two reservoirs whose electrochemical potentials are slowly modulated. 
While our system is an example of adiabatic quantum (Thouless) pumping, the explicit calculation in Sec. \ref{sec:appli} is still quasi-classical.
In the adiabatic regime, the work is expressed as the line integral of the vector potential in the parameter space.
On the other hand, the lower bound of the one-cycle entropy production can be written as the square of the thermodynamic length along the path.
These results are unchanged from those for the geometrical thermodynamic engine obtained in Ref.~\cite{Hino2021}.
We applied the formulation to the Anderson model within the wide-band approximation to obtain the explicit values of one-cycle averaged entropy production, thermodynamic length, adiabatic work, and effective efficiency.

Our future tasks are as follows:
(i) To calculate the thermodynamic metric tensor or the vector potential, we need to know the explicit form of the steady-state of the quantum master equation.
In other words, we cannot use our method in systems for which a steady solution is not explicitly known.
Since we cannot obtain non-equilibrium steady states analytically in most cases, we need to extend our formulation to cases where the steady state is not explicitly known.
(ii) 
When we consider a perfectly periodic case, the effective efficiency $\eta^{\rm eff}$ can reach unity if we choose parameters satisfying $W\ge 0$.
Of course, this setup is not a thermodynamic situation, but it might be possible to achieve this if we know the details of the properties of the engine \emph{a priori}.
The perfectly periodic situation is needed to be analyzed in detail.
(See Ref.~\cite{Yoshii22} for the analysis under a perfectly periodic control of parameters).
Indeed, it is known that the energy fluctuation under unitary operations is subject to certain bounds~\cite{Tajima18}, even though we ignore the energy fluctuations in our non-unitary dynamics.
Including the control costs in our formulation and clarifying the role of such fluctuations will be a subject of our future work.
As an alternative interpretation, an engine with efficiency equal to unity can be achieved in some setups such as the sensor-gate model~\cite{Sekimoto10} and the autonomous Maxwell's demon~\cite{Shiraishi15}.
Therefore, we need to consider the implementation of real electrochemical engines reaching $\eta^{\rm eff}\to 1$ in the future.
(iii) Because the present method, at least, for the argument after Sec. \ref{sec:linear-response},  is restricted to the adiabatic case $\epsilon\to 0$, we will need to extend the analysis to the non-adiabatic regime of finite $\epsilon$.
Reference \cite{FHHT} obtained the non-adiabatic solution of a classical master equation and a geometrical representation of the non-adiabatic current in a two-level system.
We expect to apply these methods to investigate the non-adiabatic effect in heat engines.
(iv) Although we have analyzed a quantum system, our treatment in Sec. \ref{sec:appli} is still quasi-classical. 
Thus, we have so far been unable to clarify the role of quantum coherence.
Reference \cite{Brandner-Saito} showed that quantum coherence reduces the performance of slowly driven heat engines.
On the other hand, it was shown that coherence can enhance the performance of heat engines in Ref. \cite{coherence}.
Therefore, we will need to resolve the current confusing situation in which quantum coherence leads to the enhancement or reduction of efficiency by using a fully quantum mechanical model.
(v) We assumed that the master equation \eqref{master} is still Markovian even though parameter modulation is present.
However, this assumption may to not be valid in general as indicated in Ref.~\cite{Mizuta21}. 
This is important, because the modulation process might become non-CPTP if we use the Markovian dynamics described by Eq.~\eqref{master}.
Therefore, the effect of non-Markovianity arising from the modulation will need to be clarified. 

\section*{Acknowledgements}
The authors thank Yuki Hino for fruitful discussions. 
We also appreciate useful comments by Hiroyasu Tajima, Naoto Shiraishi, Kiyoshi Kanazawa and Kaoru Mizuta. 
RY appreciates useful comments by Asahi Yamaguchi.
This work is partially supported by a Grant-in-Aid of MEXT for Scientific Research (Grant Nos. 16H04025 and 21H01006).
The work of RY is supported by JSPS Grant-in-Aid for Scientific Research (KAKENHI Grant Nos.~19K14616 and 20H01838). 

\appendix

\section{Positive semidefinite property of $S^{\rm HS}$}\label{sagawa_KL_divergence}

In this appendix, we prove the positive semidefinite property of the relative entropy
\begin{equation}\label{positive_KL}
S^{\rm HS}(\hat{\rho}||\hat{\sigma})\ge 0 
\end{equation}
based on the description in Ref.~\cite{Sagawa20}.
See also Refs.~\cite{Petz86,Petz03,Ruskai02,Hiai11}.

Now, let us prove Eq.~\eqref{positive_KL},
For this purpose, first we introduce the Hilbert-Schmidt inner product defined as
\begin{equation}
\langle \hat{Y}, \hat{X} \rangle_{\rm HS}:={\rm Tr}[\hat{Y}^\dagger \hat{X}] .
\label{Hilbert-Schmidt}
\end{equation}
We also introduce the left and right multiplications of $\hat{\rho}$ as
\begin{equation}
\mathscr{L}_{\hat{\rho}}(\hat{X}):=\hat{\rho} \hat{X}; 
\quad
\mathscr{R}_{\hat{\rho}}(\hat{X}):=\hat{X}\hat{\rho} .
\end{equation}
Here, we assume that $\mathscr{L}_{\hat{\rho}}$ and $\mathscr{R}_{\hat{\rho}}$ are commutable and Hermitians with respect to the Hilbert-Schmidt inner product.
Then, we introduce the modular operator 
\begin{equation}\label{modular}
\mathscr{D}_{\hat{\rho},\hat{\sigma}}(\hat{X}):=\mathscr{L}_{\hat{\rho}}\mathscr{R}_{\hat{\sigma}^{-1}}(\hat{X})=\hat{\rho}\hat{X}\hat{\sigma}^{-1} .
\end{equation}
Let $\hat{\rho}=\sum_i p_i \hat{P}_i$ and $\hat{\sigma}=\sum_i q_i \hat{Q}_i$ be the spectral decompositions, where $\hat{P}_i$ and $\hat{Q}_i$ are the projections onto the eigenspace with the assumption $p_i\ne p_j$ and $q_i \ne q_j$ for any pair of $i\ne  j$. 
Then, the spectral decomposition of $\mathscr{D}_{\hat{\rho},\hat{\sigma}}$ is expressed as
\begin{equation}
\mathscr{D}_{\hat{\rho},\hat{\sigma}}=\sum_{i,j}\frac{p_i}{q_j}\mathscr{P}_{ij} ,
\end{equation}
where $\mathscr{P}_{ij}$ is defined as $\mathscr{P}_{ij}:=\hat{P}_i\hat{X}\hat{Q}_j$.

Now, let us introduce a divergence-like quantity or Petz's quasi-entropy~\cite{Petz86,Petz03,Hiai11}: 
\begin{align}\label{Petz_entropy}
D^f(\hat{\rho}||\hat{\sigma}):&=\langle \hat{\sigma}^{1/2}, f(\mathscr{D}_{\hat{\rho},\hat{\sigma}})\hat{\sigma}^{1/2}\rangle_{\rm HS} 
\nonumber\\
&=\sum_{i,j} q_j f\left(\frac{p_i}{q_j}\right){\rm Tr}[\hat{P}_i\hat{Q}_j] .
\end{align}
If we use $f(x)=x \ln x$, 
with the aid of
$f(\mathscr{D}_{\hat{\rho},\hat{\sigma}})=\mathscr{L}_{\hat{\rho}}\mathscr{R}_{\hat{\sigma}^{-1}} (\ln \mathscr{L}_{\hat{\rho}}+\ln \mathscr{R}_{\hat{\sigma}^{-1}})$ and the commutability of $\mathscr{L}_{\hat{\rho}}$ and $\mathscr{R}_{\hat{\sigma}^{-1}}$,  we obtain
\begin{equation}
f(\mathscr{D}_{\hat{\rho},\hat{\sigma}})(\hat{X})=
(\hat{\rho}\ln \hat{\rho})\hat{X}\hat{\sigma}^{-1}-\hat{\rho}\hat{X}(\hat{\sigma}^{-1}\ln \hat{\sigma} ) .
\end{equation}
This leads to
\begin{align}\label{D_f=S_KL}
D^f(\hat{\rho}||\hat{\sigma})&=
{\rm Tr}[\hat{\sigma}^{1/2}(\hat{\rho}\ln\hat\hat{\rho}) \hat{\sigma}^{1/2}\hat{\sigma}^{-1}]
\notag\\
&\quad -{\rm Tr}[\hat{\sigma}^{1/2}\hat{\rho}\hat{\sigma}^{1/2}(\hat{\sigma}^{-1}\ln \hat{\sigma})]
\nonumber\\
&=S^{\rm HS}(\hat{\rho}||\hat{\sigma}) .
\end{align}
Thus, the relative entropy can be mapped onto the divergence-like quantity Eq.~\eqref{Petz_entropy}.

Then, to prove the non-negativity of Eq.~\eqref{positive_KL} is sufficient to prove~\cite{Sagawa20,Petz86,Petz03,Hiai11}
\begin{equation}\label{D_positive}
D^f(\hat{\rho}||\hat{\sigma})\ge 0 .
\end{equation}
The proof of Eq.~\eqref{D_positive} is simple as follows.
$T_{ij}:={\rm Tr}[\hat{P}_i\hat{Q}_j]$ introduced in Eq.~\eqref{Petz_entropy} is a doubly stochastic matrix, i.e. the stochastic matrix $T_{ij}$ satisfies $\sum_i T_{ij}=1$.
Suppose $h(x)$ is a convex function, $T_{ij}$ and $h(x)$ satisfy
\begin{equation}
\sum_i h\left(\frac{p_i}{q_i}\right)T_{ij}\ge h\left(\frac{p'_j}{q_j}\right) ,
\end{equation}
where $p'_j:=\sum_ip_iT_{ij}$.
Thus, we obtain
\begin{equation}
D^f(\hat{\rho}||\hat{\sigma})\ge D^h(p'||q):=\sum_i q_i h\left(\frac{p_i'}{q_i}\right) .
\end{equation}
Using the convexity of $h(x)$, 
$D^f(p'||q)$ satisfies the relation $D^f(p'||q)\ge h(\sum_i q_i (p_i/q_i))=h(1)$.
Equality, i.e. $D^f(\hat{\rho}||\hat{\sigma})=h(1)$, is held if and only if $\hat{\sigma}=\hat{\rho}$.
If we choose $h(x)=x \ln x$, which is one of convex functions with $h(1)=0$, we reach Eq.~\eqref{positive_KL}.

\section{Contributoin of the housekeeping entropy}\label{housekeeping}


The nonequilibrium system we consider is sustained by an external agent, 
which requires housekeeping heat as well as excess heat, 
although the main text only contains the description for the excess heat~\cite{sagawa,yuge2}.
From the parallel discussion to Ref.~\cite{Yoshii22}, we evaluate the housekeeping entropy production in our system in this appendix.  

As shown in Refs.\,\cite{sagawa,yuge2}, we introduce a set of counting fields $\bm{\chi}$ to calculate physical observables.
As a result, Eq.\,\eqref{master} is formally modified as 
\begin{equation}\label{counting_master}
 \frac{d}{d\theta}|\hat{\rho}(\theta,\delta,\bm{\chi})\rangle
=\epsilon^{-1}\hat{K}^{\bm{\chi}}|\hat{\rho}(\theta,\delta,\bm{\chi}) \rangle ,
\end{equation}
where the set of counting fields contains two components $\bm{\chi}=(\chi_L,\chi_R$) 
are inserted to monitor the time evolution of the housekeeping entropy production in the left and right reservoirs, respectively,  
and $|\hat{\rho}(\theta,\delta,\bm{\chi})\rangle$ and $\hat{K}^{\bm{\chi}}$ are the generalized density matrix and the evolution operator, respectively.
Since $\hat{\rho}(\theta,\delta,\bm{\chi})$ behaves as $\hat{\rho}(\theta,\delta,\bm{\chi})\sim \exp[\lambda_0(\bm{\Lambda},\bm{\chi}) \theta/\epsilon]$ for large $\theta/\epsilon$ 
with the smallest eigenvalue $\lambda_0(\bm{\Lambda},\bm{\chi})$ of $\hat{K}^{\bm{\chi}}$ 
(which is reduced to zero in the limit $\bm{\chi}\to \bm{0}$) under a fixed $\bm{\Lambda}$, the housekeeping entropy flux~\cite{sagawa,yuge2} is given by
\begin{equation}
S^{\rm hk}(\phi,\delta):=
\left.\frac{\partial \lambda_0(\bm{\Lambda}(\phi,\delta),\bm{\chi})}{\partial(i\chi_L)}\right|_{\bm{\chi}=\bm{0}}
+\left.\frac{\partial \lambda_0(\bm{\Lambda}(\phi,\delta),\bm{\chi})}{\partial(i\chi_R)}\right|_{\bm{\chi}=\bm{0}}.
\label{hosekeeping}
\end{equation}  
Note that $\chi_L$ and $\chi_R$ couple to the housekeeping entropy production in the left reservoir ($\hat{\mathfrak{S}}_L=\beta(\hat H^L-\hat{N}^L\mu^L)$) and that in the right reservoir ($\hat{\mathfrak{S}}_R=\beta(\hat H^R-\hat{N}^R \mu_R)$), respectively, where
$\hat H^\alpha$ and $\hat{N}^\alpha$ ($\alpha=L,R$) stand for the Hamiltonian and the number operator $\hat{N}^\alpha:=\sum_{k,\sigma}a_{\alpha,k,\sigma}^\dagger a_{\alpha,k,\sigma}$ of the reservoir $\alpha$, respectively. 
This housekeeping entropy flux is dominant to maintain the steady-state. 

More explicitly, $\hat{K}^{\bm{\chi}}$ in Eq.\,\eqref{counting_master} can be written as
\begin{equation}\label{K^chi}
\hat{K}^{\bm{\chi}}=\hat{K}+i\sum_{\alpha=L,R}\chi_\alpha \hat{\mathcal{K}}_\alpha+O(\chi^2). 
\end{equation} 
For the explicit calculation of Eq.\,\eqref{housekeeping}, we employ the Anderson model, as in the main text. 
In the following, we consider the generating function defined by  
\begin{equation}
\ln \mathrm{Tr}\rho(\theta,\delta,\bm{\chi})=\ln \left\langle e^{i(\chi_L\hat{\mathfrak{S}}_L(\theta)+\chi_R\hat{\mathfrak{S}}_R(\theta,\delta))} \right\rangle, 
\end{equation}
where  $\langle \cdot \rangle:={\rm Tr}[\hat{\rho} {~}\cdot ]$. 
In the present case, 
only $\hat H^{\mathrm{int}}$ in the total hamiltonian does not commute with $e^{i(\chi_L\hat{\mathfrak{S}}_L+\chi_R\hat{\mathfrak{S}}_R)}$. 
In this case, one can use the technique with that used in Ref.\,\cite{Yoshii13}, namely, the counting fields can be absorbed as the phases of the interaction part.  
\begin{align}
&e^{-i\frac{\chi}{2} \hat{\mathfrak{S}}_\alpha} V_\beta d^\dagger_\sigma 
a_{\beta,k,\sigma} e^{i\frac{\chi}{2}\hat{\mathfrak{S}}_\alpha}
=e^{i\frac{\chi}{2} \mathfrak{S}_{\alpha,k}\delta_{\alpha\beta}}
V_\beta d^\dagger_\sigma a_{\beta,k,\sigma},
\\
&e^{-i\frac{\chi}{2}\hat{\mathfrak{S}}_\alpha}V_\beta a^\dagger_{\beta,k,\sigma} d_\sigma e^{i\frac{\chi}{2}\hat{\mathfrak{S}}_\alpha}
=e^{i\frac{\chi}{2}\mathfrak{S}_{\alpha,k}\delta_{\alpha\beta}}V_\beta a^\dagger_{\beta,k,\sigma} d_\sigma, 
\label{S28}
\end{align}
where ${\mathfrak{S}}_{\alpha,k}=\beta(\epsilon_k-\mu^\alpha)$ and $\delta_{\alpha\beta}$ is Kronecker's delta. 
Then, we just need to proceed the following transposition to calculate the generating function. 
\begin{align}
&a^\dagger_{\alpha,k,\sigma}\to e^{i\frac{\chi_\alpha}{2}\mathfrak{S}_{\alpha,k}}a^\dagger_{\alpha,k,\sigma},\\
&a_{\alpha,k,\sigma}\to e^{-i\frac{\chi_\alpha}{2}\mathfrak{S}_{\alpha,k}}a_{\alpha,k,\sigma}.
\end{align}
As a consequence, the similar calculation in Ref.\,\cite{Yoshii13} yields $\hat{K}^{\bm{\chi}}$ as 
\begin{align}\label{S30}
\hat{K}^{\bm{\chi}}=
\Gamma\begin{pmatrix}
-2f^{(1)}_- & f^{(1)\bm{\chi}}_+ & f^{(1)\bm{\chi}}_+ & 0 \\
f^{(1)\bm{\chi}}_- &-f^{(0)}_+ -f^{(1)}_+  & 0 & f^{(0)\bm{\chi}}_+ \\
f^{(1)\bm{\chi}}_- & 0 & -f^{(0)}_+-f^{(1)}_+ & f^{(0)\bm{\chi}}_+ \\
0 & f^{(0)\bm{\chi}}_-  & f^{(0)\bm{\chi}}_- & -2f^{(0)}_+ \\
\end{pmatrix},
\end{align}
where $f_+^{(j){\bm{\chi}}}=e^{i\chi_L\mathfrak{S}_L^j}f_L^{(j)}(\epsilon_0)+e^{i\chi_R\mathfrak{S}_R^j}f_R^{(j)}(\epsilon_0)$ and 
$f_-^{(j){\bm{\chi}}}=e^{-i\chi_L\mathfrak{S}_L^j}[1-f_L^{(j)}(\epsilon_0)]+e^{-i\chi_R\mathfrak{S}_R^j}[1-f_R^{(j)}(\epsilon_0)]$, 
$\mathfrak{S}_\alpha^j=\beta(\epsilon_0+jU-\mu^\alpha)$. 
Substituting Eq.~\eqref{S30} into Eq.~\eqref{K^chi}, we can write $\hat{\mathcal{K}}_\alpha$ as
\begin{align}\label{S31}
\hat{\mathcal{K}}_\alpha=
\begin{pmatrix}
0 & S_\alpha^1g^1_\alpha & S_\alpha^1g^1_\alpha & 0 \\
S_\alpha^1(g^1_\alpha-1) & 0 & 0 & S_\alpha^0g^0_\alpha \\
S_\alpha^1(g^1_\alpha-1) & 0 & 0 & S_\alpha^0g^0_\alpha \\
0 & S_\alpha^0(g^0_\alpha-1) & S_\alpha^0(g^0_\alpha-1) & 0 \\
\end{pmatrix},
\end{align}
with $g^j_\alpha:=(1+e^{\beta (\epsilon_0+jU-\mu^\alpha)})^{-1}$, $S_\alpha^j=\beta(\mu^\alpha-\epsilon_0-jU)$ ($j=0,1$, $\alpha=L,R$).
Using the relation 
\begin{equation}
\lambda^{(1)}_{0,\alpha}(\bm{\Lambda}(\phi,\delta))=\langle\ell_0|\hat{\mathcal{K}}_\alpha(\bm{\Lambda}(\phi,\delta))|r_0\rangle.
\end{equation}
and the expansion $\lambda_0(\bm{\Lambda}(\phi,\delta),\bm{\chi})=i\sum_\alpha\chi_\alpha \lambda^{(1)}_{0,\alpha}(\bm{\Lambda}(\phi,\delta))+O(\chi^2)$, we obtain $\lambda_0(\bm{\Lambda},\bm{\chi})$ and $S^{\rm hk}$.

The housekeeping entropy production during one cycle $(\theta,\theta+2\pi)$ is given by
\begin{equation}\label{S_hk}
\overline{S^{\rm hk}}(\theta,\delta):=\frac{1}{2\pi}\int_{\theta}^{\theta+2\pi}d\phi S^{\rm hk}(\phi,\delta).
\end{equation}
As a result, the housekeeping entropy production during one cycle of Eq.\,(\ref{S_hk}) becomes 
\begin{equation}
\overline{S^{\rm hk}}(\theta,\delta)=\frac{1}{2\pi}\sum_{\alpha={\rm L,R}}
\int_\theta^{2\pi+\theta} d\phi \langle\ell_0|\hat{\mathcal{K}}_\alpha(\bm{\Lambda}(\phi,\delta))|r_0\rangle.
\label{S_hk2}
\end{equation} 
Although $\overline{S^{\rm hk}}(\theta,\delta)$ depends on $\theta$ because of non-periodicity of $\lambda$,
We adopt $\theta=0$ for our calculation, because the choice of $\theta$ is not crucially important as will be shown later.

\begin{figure}
\centering
\includegraphics[clip,width=8cm]{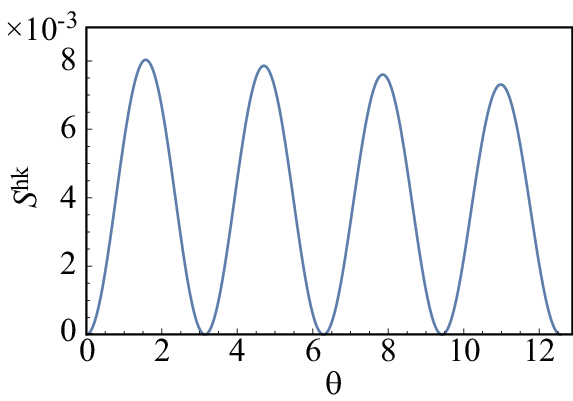}
\caption{Time evolution of  $J_{\rm hk}$.
We set $\epsilon=0.1$, $r=0.9$, $\beta U_0=0.1$, $\beta \epsilon_0=0.1$, $\beta \bar \mu =0.1$, and $\delta=\pi$.
}
\label{fig:Housekeeping_timedep}
\end{figure}

In Fig.\,\ref{fig:Housekeeping_timedep}, we plot the time dependence of the housekeeping entropy production $S^{\rm hk}$ given by Eq.\,(\ref{hosekeeping}) in the case of the Anderson model, where we set $\delta=\pi$.
We find that $S^{\rm hk}$ oscillates with time keeping its non-negativity without the quick decay from the initial condition.
This behavior supports that $S^{\rm hk}$ is insensitive to the choice of the initial condition. 
The amplitude of the oscillation of $S^{\rm hk}$ decreases with time, because of our setup $\dot{\lambda}\ge 0$ and $T\sigma\ge 0$ (see Eq.~\eqref{thermo_2nd_law}).



\begin{figure}
\centering
\includegraphics[clip,width=8cm]{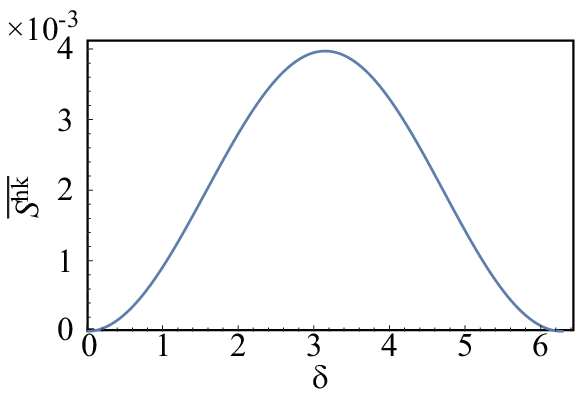}
\caption{The plot of the housekeeping entropy $S^{\rm hk}$ against $\delta$.
We set $\epsilon=0.1$, $r=0.9$, $\beta U_0=0.1$, $\beta \epsilon_0=0.1$, and $\beta \bar \mu =0.1$.
}
\label{fig:Housekeeping_deltadep}
\end{figure}

We also plot $\overline{S^{\rm hk}}$ against $\delta$ in Fig.~\ref{fig:Housekeeping_deltadep}.
This indicates that $\overline{S^{\rm hk}}$ is always non-negative and becomes zero at $\delta=0=2\pi$ and takes the maximum value at $\delta=\pi$.
This $\overline{S^{\rm hk}}$ is the housekeeping entropy in the first cycle.
The value of $\overline{S^{\rm hk}}$ is basically much larger than $2\pi \sigma$ except for the region near $\delta=0=2\pi$.
The Joule heat to maintain the steady state is $T S^{\rm hk}$.
Since there are no housekeeping entropy and Joule heating at $\delta=0=2\pi$, such contributions are not essentially important for our argument.

\section{Some detailed properties of general framework}\label{app:slow-driving}

In this appendix, we explain some general properties of the quantum master equation such as the outline of the perturbation method with a slowly modulated parameter in Appendix~\ref{slow_driving} and the mathematical description of pseudo-inverse of transition matrix in Appendix~\ref{app:pseudo-inverse}.

\subsection{Slow-driving perturbation}\label{slow_driving}

In this subsection, we explain the outline of the perturbation theory of the quantum master equation with a slowly modulated parameter $\epsilon$~\cite{slow_dynamics}. 
First, we expand the solution of Eq. (\ref{master}) in terms of $\epsilon$ as
\begin{align}\label{p-expand}
    |\hat{\rho}(\theta)\rangle = \sum_{n=0}^{\infty} \epsilon^{n} |\hat{\rho}^{(n)}(\bm{\Lambda}(\theta))\rangle  
\end{align}
with $|\hat{\rho}^{(0)}\rangle=|\hat{\rho}^{\rm ss}\rangle$ as in Eq.~\eqref{p-ad}.
Since the normalization condition ${\rm Tr}[\hat{\rho}(\theta)] =1$ holds for any $\epsilon$, 
$|\hat{\rho}^{(n)}(\bm{\Lambda}(\theta))\rangle$ satisfies
\begin{align}
    &{\rm Tr}[\hat{\rho}^{\rm ss}(\bm{\Lambda}(\theta))] = 1, \\
    &{\rm Tr}[\hat{\rho}^{(n)}(\bm{\Lambda}(\theta))] =0
\end{align}
for $n\ge 1$.
Substituting them into Eq. (\ref{master}), we obtain Eq.~\eqref{steady_condition} and
\begin{align}
    \label{pn-eq}
    &\hat{K}(\bm{\Lambda}(\theta))|\hat{\rho}^{(n)}(\bm{\Lambda}(\theta))\rangle = \frac{d}{d\theta}|\hat{\rho}^{(n-1)}(\bm{\Lambda}(\theta))\rangle 
\end{align}
for $n\ge 1$,
By using the pseudo-inverse $\hat{K}^{+}(\bm{\Lambda}(\theta))$ of $\hat{K}(\bm{\Lambda}(\theta))$, Eq. (\ref{pn-eq}) can be written as
\begin{align}\label{pn}
    |\hat{\rho}^{(n)}(\bm{\Lambda}(\theta))\rangle 
    &= \hat{K}^{+}(\bm{\Lambda}(\theta)) \frac{d}{d\theta} |\hat{\rho}^{(n-1)}(\bm{\Lambda}(\theta))\rangle \notag \\
    &= \left(\hat{K}^{+}(\bm{\Lambda}(\theta)) \frac{d}{d\theta} \right)^{n} |\hat{\rho}^{\mathrm{ss}}(\bm{\Lambda}(\theta))\rangle.
\end{align}
Ignoring terms of $O(\epsilon^{2})$ and higher in Eq. (\ref{p-expand}), we obtain Eq. (\ref{rho^1}) of the main text.

\subsection{Pseudo-inverse of the transition matrix}\label{app:pseudo-inverse}

In this subsection, we introduce the pseudo-inverse $\hat{K}^{+}(\bm{\Lambda})$ of $\hat{K}(\bm{\Lambda})$, which satisfies following conditions \cite{inverse, Mandal}
\begin{align}
    \label{c1}
    &\hat{K}^{+}(\bm{\Lambda})\hat{K}(\bm{\Lambda}) =\hat{K}(\bm{\Lambda})\hat{K}^{+}(\bm{\Lambda}) = 1 - |\hat{\rho}^{\mathrm{ss}}(\bm{\Lambda})\rangle\langle 1|  , \\
    \label{c3}
    &\hat{K}^{+}(\bm{\Lambda})|\hat{\rho}^{\mathrm{ss}}(\bm{\Lambda})\rangle = 0, \\
    \label{c4}
    &\langle 1|\hat{K}^{+}(\bm{\Lambda}) =0.
\end{align}

Here, the eigenequations are given by
\begin{align}\label{left_eigen_eq}
\langle \ell_{m}(\bm{\Lambda})| \hat{K}(\bm{\Lambda}) &=
\varepsilon_{m}(\bm{\Lambda}) \langle \ell_{m}(\bm{\Lambda})| \\
\hat{K}(\bm{\Lambda}) |r_{m}(\bm{\Lambda})\rangle &=
\varepsilon_{m}(\bm{\Lambda})  |r_{m}(\bm{\Lambda})\rangle , 
\label{right_eigen_eq}
\end{align} 
We note that $\varepsilon_{0}(\bm{\Lambda})=0$, then $|r_{0}(\bm{\Lambda})\rangle = |\hat{\rho}^{\mathrm{ss}}(\bm{\Lambda})\rangle$ and $\langle \ell_{0}(\bm{\Lambda})| = \langle 1|$.
For simplicity, we assume that these eigenstates do not degenerate throught this paper.
Thus, we need to solve the eigenvalue problem \eqref{left_eigen_eq} or Eq.~\eqref{right_eigen_eq} to express the psedo-inverse operator.

The definition of $\hat{K}^+$ in Eq.~\eqref{K+} satisfies the requirements of Eqs.\eqref{c1}-\eqref{c4}.
Indeed, using Eqs.~\eqref{K-expand} and \eqref{K+} we obtain
\begin{align}\label{proof_c1}
\hat{K}^+\hat{K}&=1
\sum_{m\ne0}\sum_n \frac{\varepsilon_n}{\varepsilon_m}
|r_m\rangle\langle \ell_m|r_n\rangle \langle \ell_n | 
\notag\\
&=\sum_{m\ne 0}\sum_{n\ne 0}\frac{\varepsilon_n}{\varepsilon_m}|r_m\rangle\langle\ell_n|\delta_{mn}
\notag\\
&=\sum_{n\ne 0} 
|r_n\rangle \langle \ell_n|+|r_0\rangle \langle \ell_0|-|\hat{\rho}^{\rm ss}\rangle \langle 1|  
\notag\\
&=1-|\hat{\rho}^{\rm ss}\rangle \langle 1| ,
\end{align}
where we have used $\varepsilon_0=0$ in the second line, $|r_0\rangle=|\hat{\rho}^{\rm ss}\rangle$ and $\langle \ell_0|=\langle 1|$ in the third line, 
and $\sum_n|r_n\rangle \langle \ell_n|=1$ in the last expression.
The second expression of Eq.~\eqref{c1} can be obtained by the parallel calculation to Eq.~\eqref{proof_c1}. 
Equation \eqref{c3} is the definition of the right zero eigenvector $|r_0\rangle=|\hat{\rho}^{\rm ss}\rangle$.
The proof of Eq. \eqref{c4} is also straightfoward.
Indeed, substituting Eq.~\eqref{K+} into the left hand side of Eq.~\eqref{c4} we can write
\begin{align}
\langle 1|\hat{K}^+&=\sum_{m\ne 0}\frac{1}{\varepsilon_m}\langle 1|r_m\rangle \langle \ell_m| 
=0,
\end{align}
where we have used the orthogonal relation $\langle \ell_0|r_m\rangle:=\langle 1|r_m\rangle=0$ for $m\ne 0$.
Thus, Eq.~\eqref{K+} satisfies all requirements of the pseudo-inverse.

With the aid of Eq.~\eqref{c4} it is straightfoward to obtain
\begin{equation}\label{Tr[K+A]=0}
{\rm Tr}[\hat{K}^+\hat{A}]=\langle 1|\hat{K}^+\hat{A}|1\rangle=0
\end{equation}
for an arbitrary matrix $\hat{A}$.
Thus, we have the relation
\begin{equation}\label{Tr[D_mu]=0}
{\rm Tr}[\partial_\mu\partial_\nu \hat{\rho}^{\rm ss}]=
{\rm Tr}\left[\hat{K}^+\frac{\partial}{\partial \Lambda_\mu}
\left(\hat{K}^+\frac{\partial}{\partial \Lambda_\nu}\right) \hat{\rho}^{\rm ss}
\right]=0 .
\end{equation}
Multiplying $\hat{\rho}^{\rm ss}$ on the both side of the relation
\begin{equation}
\partial_\mu\partial_\nu\ln\hat{\rho}^{\rm ss}
=(\hat{\rho}^{\rm ss})^{-1}\partial_\mu\partial_\nu\hat{\rho}^{\rm ss}
-\partial_\mu \ln \hat{\rho}^{\rm ss}\cdot \partial_\nu \ln\hat{\rho}^{\rm ss}
\end{equation}
and take the trace with the aid of Eq.~\eqref{Tr[D_mu]=0} we obtain
\begin{equation}\label{Fisher=Hessian}
{\rm Tr}[\hat{\rho}^{\rm ss}\partial_\mu\partial_\nu \ln \hat{\rho}^{\rm ss}]
=-{\rm Tr}[\hat{\rho}^{\rm ss}\partial_\mu \ln \hat{\rho}^{\rm ss} \cdot \partial_\nu \ln \hat{\rho}^{\rm ss}] .
\end{equation}

\section{The derivation of the transition matrix}\label{K-matrix}


In this appendix, we present the details of the derivation of the transition matrix which appears in Eq.~\eqref{master}.
This appendix consists of two part.
In the first part, we explain the outline of the derivation of the quantum master equation.
In the second part, we derive the transition matrix in Eq.~\eqref{master}. 
Note that the most descriptions in this appendix are only applicable to the Anderson model used in Sec.~\ref{sec:appli}.

\subsection{Quantum master equation for the Anderson model}

In this subsection we explain the outline of the derivation of the quantum master equation for the Anderson model used in Sec.~\ref{sec:appli}. 
In this part, we use the time $t$ instead of using the phase $\theta$ because the system may not relax to a steady state if the time is not long enough. 
Before moving to the detail, we mention that the double bracket notation is used for the super-vector. 
The super-vectors appearing in this part corresponds to those in the main text as follows. 
\begin{align}
&|e,e\rangle\rangle= \left(\begin{array}{c} 1\\0\\0\\0\end{array}\right),\ 
|\uparrow, \uparrow\rangle\rangle=\left(\begin{array}{c} 0\\1\\0\\0\end{array}\right),\ \nonumber\\
&|\downarrow, \downarrow\rangle\rangle=\left(\begin{array}{c} 0\\0\\1\\0\end{array}\right),\ 
|d, d\rangle\rangle=\left(\begin{array}{c} 0\\0\\0\\1\end{array}\right). 
\end{align}

As is assumed, the density matrix of the total system $\hat{\rho}^{\rm tot}$ is decomposed into the matrix of system $\hat{\rho}$ and 
the matrix of the bath in thermal equilibrium $\hat{\rho}^{\textrm{r}}$ at initial time $t_0$, 
\begin{equation}
\hat{\rho}^{\rm tot}(t_0)=\hat{\rho}(t_0)\otimes\hat{\rho}^{\mathrm{r}}.
\label{initrho}
\end{equation}
The time evolution of $\hat{\rho}^{\rm tot}(t)$ is described by Liouvillian as 
\begin{equation}
\frac{d}{dt}\hat{\rho}^{\rm tot}(t)=\hat{\mathcal{K}}^{\rm tot}\hat{\rho}^{\rm tot}(t). 
\label{QME0}
\end{equation}
As is the case of Hamiltonian (\ref{H_s}), the Liouvillian $\hat{\mathcal{K}}^{\rm tot}$ for the total system can be decomposed into $\hat{\mathcal{K}}$, $\hat{\mathcal{K}}^{\mathrm{r}}$, and $\hat{\mathcal{K}}^{\mathrm{int}}$. 
As can be seen from the form of time evolution equation, Dyson's equation in quantum system can be used. 
By the Laplace transformation of $\hat{\rho}^{\rm tot}(t)$ 
\begin{equation}
\hat{\varrho}^{\rm tot}(z)=\int^\infty_{t_0}dt e^{-z(t-t_0)}\hat{\rho}^{\rm tot}(t),
\end{equation}
Eq.\ (\ref{QME0}) becomes 
\begin{align}
\hat{\varrho}^{\rm tot}(z)&=\frac{1}{z-\hat{\mathcal{K}}^{\rm tot}}\hat{\rho}(t_0)
\notag\\
&=
(G^0(z)+G^0(z)\hat{\mathcal{K}}_{\mathrm{int}}G^0(z)+\cdots)\hat{\rho}^{\rm tot}(t_0),
\label{QMELspace}
\end{align}
where $G^0(z)=(z-\hat{\mathcal{K}}-\hat{\mathcal{K}}^{\mathrm{r}})^{-1}$. 
From Eqs.\ (\ref{initrho}) and (\ref{QMELspace}), 
the reduced density matrix which is obtained by tracing out the bath degrees of freedom becomes 
\begin{equation}
\hat{\varrho}(z)=\mathrm{Tr}_{\mathrm{r}}\left[\left(G^0(z)+
G^0(z)\hat{\mathcal{K}}^{\mathrm{int}}G^0(z)\hat{\mathcal{K}}^{\mathrm{int}}G^0(z)\right)\hat{\rho}(t_0)\otimes \hat{\rho}^{\mathrm{r}}\right],
\label{rhosz}
\end{equation}
in the second order of $\hat{\mathcal{K}}^{\mathrm{int}}$. 
It can be shown that the term linear to $\hat{\mathcal{K}}^{\mathrm{int}}$ vanishes. 
By using $\hat{\mathcal{K}}^{\mathrm{r}}\hat{\rho}^{\mathrm{r}}=0$, the first term of RHS of Eq.\ (\ref{rhosz}) can be rewritten as 
\begin{equation}
\mathrm{Tr}_{\mathrm{r}}\left[G^0(z)\hat{\rho}(t_0)\otimes \hat{\rho}^{\mathrm{r}}\right]=
G^0_{\mathrm{s}}(z)\hat{\rho}(t_0),
\label{Tr1st}
\end{equation}
where $G^0_{\mathrm{s}}(z)=(z-\hat{\mathcal{K}})^{-1}$. 
Now we define the effective Liouvillian $\hat{\mathcal{K}}_\mathrm{eff}$, 
which describes the time evolution of $\hat{\rho}$ as 
\begin{equation}
\hat{\varrho}(z)=\frac{1}{z-\hat{\mathcal{K}}^\mathrm{eff}(z)}\hat{\rho}(t_0).
\label{rho_sTE}
\end{equation}
This is equivalent to the following time evolution equation 
\begin{equation}
\frac{d}{dt}\hat{\rho}(t)=\int_{t_0}^td\tau \hat{\mathcal{K}}^{\mathrm{eff}}(t-\tau)\hat{\rho}(\tau).
\label{terhoeff}
\end{equation}
Here we can see the non-Markovian memory effect. 
By decomposing $\hat{\mathcal{K}}^{\mathrm{eff}}(z)$ into the free part $\hat{\mathcal{K}}$ and the ``self energy part" $\Sigma$ as 
$\hat{\mathcal{K}}^{\mathrm{eff}}=\hat{\mathcal{K}}+\hat{\Sigma}(z) $, 
it becomes more clear that the memory effect is induced by the interaction. 
By expanding (\ref{rho_sTE}) in $\mathcal{K}_\mathrm{eff}(z)$, we obtain 
\begin{equation}
\hat{\varrho}(z)=\left(G^0_{\mathrm{s}}(z)+
G^0_{\mathrm{s}}(z)\hat{\Sigma}(z) G^0_{\mathrm{s}}(z)+\cdots \right)\hat{\rho} (t_0) .
\label{rhoeffz}
\end{equation}
From Eqs.\ (\ref{rhosz}), (\ref{Tr1st}), and (\ref{rhoeffz}), we can easily see that the second order term in Eq.\ (\ref{rhosz}) 
is equal to second term in Eq.\ (\ref{rhoeffz}). 
Thus the lengthy calculation yields\cite{Leijnse} 
\begin{equation}
\hat{\Sigma}(z)=- \sum_{c,c^\prime,\xi=\pm } \sum_{\sigma=\uparrow,\downarrow}c c^{\prime} J^{c^\prime}_{-\xi, \sigma}  
|aa^\prime\rangle \rangle \langle\langle aa^\prime|
 J^{c}_{\xi, \sigma} I(\xi, c, a, a^\prime), 
\end{equation}
where $|a, b \rangle\rangle =|a\rangle \langle b| $ is the two state vector and $J^{c}_{\xi, \uparrow}$ is the ladder operators (for the case of $\downarrow$, the definition is the same) defined as 
\begin{eqnarray}\label{J_+^+up}
&&J^{+}_{+, \uparrow}=\sum_{a=e,\uparrow,\downarrow,d}\left(|e, a \rangle\rangle\langle\langle \uparrow, a | 
+ |\downarrow, a \rangle\rangle\langle\langle d, a |\right), \\
&&J^{+}_{-, \uparrow}=\sum_{a=e,\uparrow,\downarrow,d}\left(|\uparrow, a \rangle\rangle\langle\langle e, a | 
+ |d, a \rangle\rangle\langle\langle \downarrow, a |\right), \\
&&J^{-}_{+, \uparrow}=\sum_{a=e,\uparrow,\downarrow,d}\left(|a, \uparrow \rangle\rangle\langle\langle a, e | 
+ |a, d \rangle\rangle\langle\langle a, \downarrow |\right), \\
&&J^{-}_{-, \uparrow}=\sum_{a=e,\uparrow,\downarrow,d}\left(|a, e \rangle\rangle\langle\langle a, \uparrow | 
+ |a, \downarrow \rangle\rangle\langle\langle a, d |\right), 
\label{J_-^-up}
\end{eqnarray}
and $I(\xi, c, a, a^\prime)$ is given by 
\begin{equation}
I=V^2\sum_k \frac{f^{-\xi c} (\omega_k)}{z+i\xi \omega_{k}+i\Delta_{a,a^\prime}}, 
\label{spectrum}
\end{equation}
where $\Delta_{a,a^\prime}:=\epsilon_a-\epsilon_{a^\prime}$ is the energy difference between the state $a$ and $a'$. 
Here we define $f^{-\xi c}_L+f^{-\xi c}_R=f^{-\xi c}$.
In the case of flat band $d \omega_k /d k=\mathrm{const}$, the function $I$ which describes the effect of spectrum in the reservoir on tunneling process, can be rewritten as 
\begin{equation}
I=\frac{\Gamma}{\pi}  \int ^{D}_{-D} d\omega \frac{f^{-\xi c} (\omega)}{z+i\xi \omega+i\Delta_{a,a^\prime}},
\end{equation}
where the line width $\Gamma$ is defined as $\Gamma =\pi \mathfrak{n} V^2$ with the density of states $\mathfrak{n}$ in reservoirs. 

Next we make an assumption which corresponds to neglect the memory effect in Eq.\ (\ref{terhoeff}). 
This is valid when the time scale of the dynamics of the system is much larger than that of bath. 
By taking the long-time limit $z\rightarrow +0$, 
we can use the Sokhotski-Plemelj relation $\lim_{\eta\rightarrow +0} (\omega+i\eta)^{-1}=-i\pi \delta(\omega)+P \omega^{-1}$. 
Assuming the wide band limit $D\rightarrow \infty$, the imaginary part of $\Sigma$ can be negligible and thus, we obtain 
\begin{equation}
\hat{\Sigma}(+0)=-\frac{\Gamma }{2} \sum_{c,c^\prime,\xi,\sigma } c c^{\prime} J^{c^\prime}_{-\xi, \sigma}  |a a^\prime \rangle \rangle \langle\langle aa^\prime|
 J^{c}_{\xi, \sigma} f_\alpha^{-c\xi }(-\xi\Delta_{aa^\prime}). 
\end{equation}
By replacing $\hat{\Sigma}(z)\rightarrow \hat{\Sigma}(+0)$ (long-time limit),  Eq.\ (\ref{terhoeff}) can be rewritten as
\begin{equation}
\frac{d}{dt}\hat{\rho}(t)=\hat{\mathcal{K}}^{\mathrm{eff}}(z=+0) \hat{\rho}(t).
\end{equation}
By using this method, we arrive at Eq.\ (\ref{master}) by setting $\hat{\mathcal{K}}^{\rm eff}(z=+0)=2\pi \Gamma \hat{K}$ and $\theta=2\pi (t-t_0)/\tau_p$ where $t_0$ and $\tau_p$ are the time to reach the periodic state and the period, respectively.

\subsection{The explicit expression of the transition matrix $\hat{K}$ in Eq.~\eqref{master} for the Anderson model}

As is explained in the previous subsection, the transition matrix $\hat{\mathcal{K}}=\hat{\mathcal{K}}_{\rm eff}$ is given by
\begin{equation}
\hat{\mathcal{K}}
=-\frac{\Gamma}{2} \sum cc^\prime J^{c^\prime}_{-\xi, \sigma} |aa^\prime\rangle\rangle\langle\langle aa^\prime|J^{c}_{\xi, \sigma}f_\alpha^{-c\xi}(-\xi \Delta_{aa^\prime}),
\end{equation}
where $\alpha=$L or R, $a$ and ,$a^\prime$ can take the state of  $e$, $\uparrow$,$\downarrow$ or $d$, $c=\pm 1$, $\xi=\pm 1$, and $\sigma$ is the spin taking the values $\sigma =\uparrow$ and $\downarrow$. 
All the subscriptions and superscriptions are summed up. 
Here $J_{\xi,\sigma}^c$ is the ladder operator given by 
Eqs.~\eqref{J_+^+up}-\eqref{J_-^-up}.

\begin{widetext}

\subsubsection{Calculation for $\langle\langle i,i|\hat{\mathcal{K}}|j, j\rangle\rangle$}

In this part, we write the result of operation $\hat{\mathcal{K}}$ onto the super-vector $|e,e\rangle\rangle$,  $|\uparrow,\uparrow\rangle\rangle$, 
$|\downarrow,\downarrow\rangle\rangle$, and $|d,d\rangle\rangle$ as 
\begin{align}
\hat{\mathcal{K}}|e, e\rangle\rangle
=&-2\Gamma  f_{+}(\epsilon_0)|e ,e\rangle \rangle +\Gamma f_{+}(\epsilon_0)|\uparrow ,\uparrow\rangle \rangle
+\Gamma f_{+}(\epsilon_0)|\downarrow ,\downarrow\rangle \rangle, \\
\hat{\mathcal{K}}|\uparrow, \uparrow\rangle\rangle
=&\Gamma f_-(\epsilon_0)|e,e\rangle\rangle+\Gamma f_+(\epsilon_0+U) |d,d\rangle \rangle -\Gamma [f_-(\epsilon_0)+f_+(\epsilon_0+U)] |\uparrow ,\uparrow\rangle \rangle,\\ \hat{\mathcal{K}}|\downarrow, \downarrow\rangle\rangle
=&\Gamma f_-(\epsilon_0)|e,e\rangle\rangle+\Gamma f_+(\epsilon_0+U) |d,d\rangle \rangle -\Gamma [f_-(\epsilon_0)+f_+(\epsilon_0+U)] |\downarrow ,\downarrow\rangle \rangle,\\ \hat{\mathcal{K}}|d, d\rangle\rangle
=&-2\Gamma  f_-(\epsilon_0+U)|d ,d\rangle \rangle +\Gamma f_-(\epsilon_0+U)|\uparrow ,\uparrow\rangle \rangle
+\Gamma f_-(\epsilon_0+U)|\downarrow ,\downarrow\rangle \rangle, 
\end{align}
where we use $\Delta_{a,a^\prime}=-\Delta_{a^\prime, a}=\epsilon_0$. 
These results yields the each components of K-matrix as follows:
\begin{align}
&\left(\begin{array}{cccc}
\langle\langle d, d|\hat{\mathcal{K}}|d,d\rangle\rangle 
& \langle\langle d, d|\hat{\mathcal{K}}|\uparrow,\uparrow\rangle\rangle 
& \langle\langle d, d|\hat{\mathcal{K}}|\downarrow,\downarrow\rangle\rangle
& \langle\langle d, d|\hat{\mathcal{K}}|e,e\rangle\rangle\\
\langle\langle \uparrow, \uparrow|\hat{\mathcal{K}}|d,d\rangle\rangle 
& \langle\langle \uparrow, \uparrow|\hat{\mathcal{K}}|\uparrow,\uparrow\rangle\rangle 
& \langle\langle \uparrow, \uparrow|\hat{\mathcal{K}}|\downarrow,\downarrow\rangle\rangle
& \langle \langle\uparrow, \uparrow|\hat{\mathcal{K}}|e,e\rangle\rangle\\
\langle\langle \downarrow, \downarrow|\hat{\mathcal{K}}|d,d\rangle\rangle 
& \langle\langle \downarrow, \downarrow|\hat{\mathcal{K}}|\uparrow,\uparrow\rangle\rangle 
& \langle\langle \downarrow, \downarrow|\hat{\mathcal{K}}|\downarrow,\downarrow\rangle\rangle
& \langle\langle \downarrow, \downarrow|\hat{\mathcal{K}}|e,e\rangle\rangle\\
\langle\langle e, e|\hat{\mathcal{K}}|d,d\rangle\rangle 
& \langle\langle e, e|\hat{\mathcal{K}}|\uparrow,\uparrow\rangle\rangle 
& \langle\langle e, e|\hat{\mathcal{K}}|\downarrow,\downarrow\rangle\rangle
& \langle\langle e, e|\hat{\mathcal{K}}|e,e\rangle\rangle\\
\end{array}\right)\nonumber\\
&=\Gamma\left(\begin{array}{cccc}
-2f_-(\epsilon_0+U)
& f_+(\epsilon_0+U)
& f_+(\epsilon_0+U)
& 0\\
f_-(\epsilon_0+U)
& -f_-(\epsilon_0)-f_+(\epsilon_0+U)
& 0
& f_{+}(\epsilon_0)\\
f_-(\epsilon_0+U)
& 0
& -f_-(\epsilon_0)-f_+(\epsilon_0+U)
& f_{+}(\epsilon_0)\\
0
& f_-(\epsilon_0)
& f_-(\epsilon_0)
& -2 f_+(\epsilon_0)
\end{array}\right)\nonumber\\
&=\Gamma\left(\begin{array}{cccc}
-2f_-^{(1)}
& f_+^{(1)}
& f_+^{(1)}
& 0\\
f_-^{(1)}
& -f_-^{(0)}-f_+^{(1)}
& 0
& f_+^{(0)}\\
f_-^{(1)}
& 0
& -f_-^{(0)}-f_+^{(1)}
& f_+^{(0)}\\
0
& f_-^{(0)}
& f_-^{(0)}
& -2 f_+^{(0)}
\end{array}\right).
\end{align}
This is identical to Eq.~\eqref{Ville_9_0806}.

\end{widetext}

\section{Explicit expression of $\hat{K}^+$ for the Anderson model }\label{sec:K^+}

\begin{widetext}
In this appendix we explicitly write the expression of $\hat{K}^+$ for the Anderson model, which is needed in the explicit calculation of the density matrix.
Substituting Eqs.~\eqref{1_eigen}, \eqref{2_eigen}, \eqref{3_eigen}, \eqref{left_1}, \eqref{right_1}, \eqref{left_2}, \eqref{right_2}, \eqref{left_3}, \eqref{right_3} into Eq.~\eqref{K+} we obtain
\begin{equation}\label{K+_exp}
\hat{K}^+=
\mathcal{C}\begin{pmatrix}
f_-^{(1)}a_1 & f_+^{(1)}a_2 & f_+^{(1)}a_2 & f_+^{(0)}f_+^{(1)}a_3\\
f_-^{(1)}a_2 & a_4 & a_5 &  f_+^{(0)}a_6 \\
f_-^{(1)}a_2 & a_5 & a_4 & f_+^{(0)}a_6 \\
f_-^{(0)}f_-^{(1)}a_3 & f_-^{(0)}a_6 & f_-^{(0)}a_6 & f_+^{(0)}a_1 \\
\end{pmatrix} ,
\end{equation}
where
\begin{align}
\mathcal{C}&=\frac{1}{8(f_+^{(0)}+f_-^{(1)})^2(f_-^{(0)}+f_+^{(1)})}, \\
a_1&=-f_-^{(0)}(f_+^{(0)}+f_-^{(1)})^2-16f_+^{(1)} , \\
a_2&=f_-^{(0)}(f_+^{(0)}+f_-^{(1)})^2+8(f_+^{(0)}-f_-^{(1)}) , \\
a_3&=16-(f_+^{(0)}+f_-^{(1)})^2 , \\
a_4&=-8[(f_+^{(0)})^2+(f_-^{(1)})^2]-f_-^{(0)}f_+^{(1)}(f_+^{(0)}+f_-^{(1)})^2 , \\
a_5&=16f_-^{(1)}f_+^{(0)}-f_-^{(0)}f_+^{(1)}(f_-^{(1)}+f_+^{(0)}) , \\
a_6&=8(f_-^{(1)}-f_+^{(0)})+f_+^{(1)}(f_-^{(1)}+f_+^{(0)})^2.
\end{align}
\end{widetext}
We can easily check that the form of the pseudo-inverse in Eq. (\ref{K+}) satisfies the above conditions (\ref{c1}) - (\ref{c4}).


\section{Some detailed properties for the Anderson model}\label{app:detailed_Anderson}

In this appendix, we present detailed properties of the Anderson model.
In particular, we write the explicit form of $\mathcal{A}_\mu$.

The system Hamiltonian is expressed as the matrix form:
\begin{equation}\label{H_s:matrix}
\hat{H}=\begin{pmatrix}
U(\theta) & 0 & 0 & 0 \\
0 & \epsilon_0 & 0 & 0 \\
0 & 0 & \epsilon_0 & 0\\
0 & 0 & 0 & 0 \\
\end{pmatrix} .
\end{equation}
With the aid of Eqs.~\eqref{U(theta)}, \eqref{density_matrix} and \eqref{H_s:matrix} we obtain
\begin{equation}\label{D2}
\mathcal{F}:={\rm Tr}\left[ \frac{\partial \hat{H}}{\partial \lambda}\hat{\rho}^{\rm ss}\right]
=U_0\rho_d .
\end{equation}
Substituting the first component of \eqref{right_zero} into Eq.~\eqref{D2} we can rewrite it as
\begin{align}\label{F3}
\mathcal{F}=\frac{ U_0 f_+^{(0)}f_+^{(1)} }{2[f_+^{(0)}+f_-^{(1)}]} .
\end{align}
Note that $\mathcal{F}$ does not depend on $f^{(1)}$ explicitly.
Therefore we can write each component of the vector potential $\mathcal{A}_\mu$ in Eq.~\eqref{vector-potential} as

\begin{widetext}
\begin{align}
\label{A^0}
\mathcal{A}_0&= \lambda\left(\frac{\partial\mathcal{F}}{\partial f_+^{(1)}}-\frac{\partial\mathcal{F}}{\partial f_-^{(1)}}\right) 
\frac{\partial f_+^{(1)}}{\partial \lambda} ,\nonumber\\
&=
-\beta U_0^2\lambda(\theta)e^{\beta(\epsilon_0+U_0\lambda(\theta))}
\left\{ 
\frac{2+f_+^{(0)}}
{(2+f_+^{(0)}-f_+^{(1)})^2}
\right\}
\left[
\frac{e^{-\beta \mu^{\rm L}}}{(1+e^{\beta(\epsilon_0+U_0\lambda(\theta)-\mu^{\rm L}(\theta))})^2}
+\frac{e^{-\beta \mu^{\rm R}}}{(1+e^{\beta(\epsilon_0+U_0\lambda(\theta)-\mu^{\rm R}(\theta))})^2}
\right] ,
\\
\label{A^1}
\mathcal{A}_1&=
\lambda \overline{\mu}
\left[
\left(\frac{\partial\mathcal{F}}{\partial f_+^{(1)}}
-\frac{\partial\mathcal{F}}{\partial f_-^{(1)}}\right)\frac{\partial f_+^{(1)}}{\partial \mu^{\rm L}}
+ \frac{\partial\mathcal{F}}{\partial f_+^{(0)}}\frac{\partial f_+^{(0)}}{\partial \mu^{\rm L}} 
\right]
\nonumber\\
&=
\frac{\beta \overline{\mu}^2U_0 \lambda(\theta)}{(2+f_+^{(0)}-f_+^{(1)})^2}
\left[
\frac{f_+^{(0)}e^{\beta(\epsilon_0+U_0\lambda-\mu^{\rm L})}
\{2+f_+^{(0)} \} }
{\{
1+e^{\beta(\epsilon_0+U_0\lambda-\mu^{\rm L})} 
\}^2
}
+\frac{
(2-f_+^{(1)})f_+^{(1)}
e^{\beta(\epsilon_0-\mu^{\rm L})}
}{
\{1+e^{\beta(\epsilon_0-\mu^{\rm L})} \}^2
}
\right]
 ,
\\
\label{A^2}
\mathcal{A}_2&=\lambda \overline{\mu}
\left[
\left(\frac{\partial\mathcal{F}}{\partial f_+^{(1)}}
-\frac{\partial\mathcal{F}}{\partial f_-^{(1)}}\right)\frac{\partial f_+^{(1)}}{\partial \mu^{\rm R}}
+ \frac{\partial\mathcal{F}}{\partial f_+^{(0)}}\frac{\partial f_+^{(0)}}{\partial \mu^{\rm R}} 
\right] 
\nonumber\\
&=
\frac{\beta \overline{\mu}^2 U_0 \lambda(\theta)}{(2+f_+^{(0)}-f_+^{(1)})^2}
\left[
\frac{f_+^{(0)}e^{\beta(\epsilon_0+U_0\lambda-\mu^{\rm R})}
(2+f_+^{(0)}) }
{\{
1+e^{\beta(\epsilon_0+U_0\lambda-\mu^{\rm R})} 
\}^2
}
+\frac{
(2-f_+^{(1)})f_+^{(1)}
e^{\beta(\epsilon_0-\mu^{\rm R})}
}{
\{1+e^{\beta(\epsilon_0-\mu^{\rm R})} \}^2
}
\right]
\end{align}
where we have used
\begin{align}
\frac{\partial f_+^{(1)}}{\partial \lambda}
&=-\beta U_0e^{\beta(\epsilon_0+U_0\lambda(\theta))}
\left[
\frac{e^{-\beta \mu^{\rm L}}}{(1+e^{\beta(\epsilon_0+U_0\lambda(\theta)-\mu^{\rm L}(\theta))})^2}
+\frac{e^{-\beta \mu^{\rm R}}}{(1+e^{\beta(\epsilon_0+U_0\lambda(\theta)-\mu^{\rm R}(\theta))})^2}
\right] ,
\end{align}
\end{widetext}
\begin{align}
\frac{\partial f_+^{(0)}}{\partial \mu^\alpha}&=
\frac{
\beta \overline{\mu} e^{\beta(\epsilon_0-\mu^\alpha(\theta))}
}{
\{1+e^{\beta(\epsilon_0-\mu^\alpha(\theta))} \}^2
} ,
\\
\frac{\partial f_+^{(1)}}{\partial \mu^\alpha}&=
\frac{
\beta \overline{\mu} e^{\beta(\epsilon_0+U_0\lambda(\theta)-\mu^\alpha(\theta))}
}{
\{1+e^{\beta(\epsilon_0+U_0\lambda(\theta)-\mu^\alpha(\theta))} \}^2
} ,
\\
\frac{\partial \mathcal{F}}{\partial f_+^{(0)}}&=
\frac{(2-f_+^{(1)})U_0f_+^{(1)}}{2(2+f_+^{(0)}-f_+^{(1)})^2} ,
\\
\frac{
\partial \mathcal{F}}
{\partial f_+^{(1)}
}&=-
\frac{\partial \mathcal{F}}{\partial f_-^{(1)}}
=\frac{U_0f_+^{(0)}}{2}
\frac{2+f_+^{(0)}}
{(2+f_+^{(0)}-f_+^{(1)})^2}.
\end{align}
Substituting Eqs.~\eqref{A^0}, \eqref{A^1} and \eqref{A^2} into Eq.~\eqref{W-ad} we obtain the expression the work.


\section{Perturbation calculation of Anderson model
in $\beta U_0\ll 1$ limit }\label{app:perturbed_Anderson}

\subsection{Framework of the perturbation method}

Now, let us solve eigenvalue problem given by Eqs. \eqref{left_eigen_eq} or \eqref{right_eigen_eq} in the high temerature (or classical) limit.
In this case, $\hat{K}(\bm{\Lambda}(\theta))$ in Eq.~\eqref{Ville_9_0806} can be expanded in terms of $\tilde{\beta}:=\beta U_0$ as
\begin{equation}\label{hat_K}
\hat{K}=\hat{K}^{(0)}+\tilde{\beta} \hat{K}^{(1)} +O(\tilde{\beta}^2) ,
\end{equation}
where 
$\hat{K}^{(0)}:=\hat{K}_{\tilde{\beta}=0}$ is given by 
\begin{align}
\hat{K}^{(0)}&=
 \left(
\begin{array}{cccc}
-2g_+ & g_+ & g_+ & 0 \\
g_- & -g_+-g_-& 0 & g_+ \\
g_- & 0 & -g_+-g_- & g_+ \\
0 & g_- & g_- & -2g_+
\end{array}
\right)
\nonumber\\
&
=\left(
\begin{array}{cccc}
-2g_+ & g_+ & g_+ & 0 \\
g_- & -2 & 0 & g_+ \\
g_- & 0 & -2  & g_+ \\
0 & g_- & g_- & -2g_+
\end{array}
\right)
\end{align}
where $g_\pm :=f_\pm^{(0)}=\lim_{\tilde{\beta}\to 0}f_\pm^{(1)}$.
$\hat{K}^{(1)}$ in Eq.~\eqref{hat_K} is given by
\begin{equation}\label{K_1}
\hat{K}^{(1)}:=
\left.\frac{\partial \hat{K}}{\partial \tilde{\beta}}\right|_{\tilde{\beta}=0}=
\left(
\begin{array}{cccc}
-2\mathring{g}_- &\mathring{g}_+&\mathring{g}_+&0\\
\mathring{g}_-&-\mathring{g}_+ &0&0\\
\mathring{g}_-&0&-\mathring{g}_+ &0\\
0&0&0&0
\end{array}
\right),
\end{equation}
where $\mathring{g}_\pm:=(\partial f^{(1)}_\pm /\partial \tilde{\beta})_{\tilde{\beta}=0}$.

Then, we can adopt the perturbation method to solve the eigenvalue problem Eqs.~\eqref{left_eigen_eq} or \eqref{right_eigen_eq}. 
The eigenequation Eqs. \eqref{left_eigen_eq} and \eqref{right_eigen_eq} can be rewritten as
\begin{align}
\langle \ell_m|\hat{K}&= \left(\langle \ell^{(0)}_{m}|+\tilde{\beta} \langle \ell^{(1)}_{m}|\right)(\hat{K}^{(0)}+\tilde{\beta}\hat{K}^{(1)}) ,
\\
\hat{K} |r_{m}\rangle&=\left(\hat{K}^{(0)}+\tilde{\beta} \hat{K}^{(1)}\right)
\left(|r_{m}^{(0)}\rangle+\tilde{\beta} |r_{m}^{(1)}\rangle \right),\label{lhs}\\
\varepsilon_{m} |r_{m}\rangle &=\left( \varepsilon_{m}^{(0)}+ \tilde{\beta} \varepsilon_{m}^{(1)} \right) 
\left(|r^{(0)}_{m}\rangle+\tilde{\beta} |r_{m}^{(1)}\rangle \right) .
\label{rhs}
\end{align}

\subsection{Properties of the unperturbed state}

Now, let us develop the perturbation of the eigenvalue equations Eqs.~\eqref{left_eigen_eq} and \eqref{right_eigen_eq} for $\tilde{\beta} \ll 1$. 

Since $\hat{K}^{(0)}$ can be decomposed into the spin $\uparrow$ space and spin $\downarrow$ space 
\begin{align}
\hat{K}^{(0)}&=\hat{K}^{(0)}_\uparrow+\hat{K}^{(0)}_\downarrow 
\end{align}
with
\begin{align}
\hat{K}^{(0)}_\uparrow&:=
\left(
\begin{array}{cccc}
-g_- &0&g_+&0\\
0&-g_- &0&g_+\\
g_-&0&-g_+&0\\
0&g_-&0&-g_+
\end{array}
\right)
\notag \\
&=
\hat{\mathcal{K}}_0
\otimes
\begin{pmatrix}
1 & 0\\
0 & 1\\
\end{pmatrix}
\end{align}
and
\begin{align}
\hat{K}^{(0)}_\downarrow
&:=
\left(
\begin{array}{cccc}
-g_- &g_+&0&0\\
g_-&-g_+ &0&0\\
0&0&-g_-&g_+\\
0&0&g_-&-g_+
\end{array}
\right).
\notag \\
&=
\begin{pmatrix}
1 & 0 \\
0 & 1\\
\end{pmatrix}
\otimes
\mathcal{K}_0
,
\end{align}
where we have introduced
\begin{equation}
\hat{\mathcal{K}}_0:=
\begin{pmatrix}
-g_- & g_+ \\
g_- & -g_+ \\
\end{pmatrix}
,
\end{equation}
the properties of unperturbed state are determined by the eigenvalue problem of $\mathcal{K}_0$.
 
Their corresponding eigenvalues are, respectively, given by
\begin{equation}\label{eigenvalues_01}
\mathcal{E}_0=0, \quad \mathcal{E}_1=-2,
\end{equation}
It should be noted that the left and right eigenvectors $\langle \ell_0|$ and $|r_{0}\rangle$ corresponding to $\varepsilon_0=0$ are, respectively, given by
\begin{equation}
\langle 0| =(1,1), \quad 
|0\rangle =\frac{1}{2}
\begin{pmatrix}
g_+\\ g_-\\ 
\end{pmatrix} .
\label{rho22}
\end{equation}
The left and right eigenvectors corresponding to $\varepsilon_1=-2$ are given by
\begin{equation}\label{rho_1}
\langle {1}|=(g_-,-g_+), \quad
|{1}\rangle =
\frac{1}{2}
\begin{pmatrix}
1 \\ -1
\end{pmatrix}
 ,
\end{equation}
respectively.
It is easy to check the orthonormal relation $\langle i |j\rangle=\delta_{ij}$ for $i,j=0$ and 1.
It is also easy to check the completeness:
\begin{align}
|0\rangle\langle 0|+|1\rangle\langle 1|&=
\frac{1}{2}
\begin{pmatrix} g_+\\ g_-\\ \end{pmatrix}(1,1)
+\frac{1}{2}\begin{pmatrix}1\\ -1\\ \end{pmatrix}(g_-,-g_+) \notag\\
&=\begin{pmatrix} 1 & 0 \\ 0 & 1 \\ \end{pmatrix} ,
\end{align}
where we have used Eq.~\eqref{sum_rule}.

The full eigenstates of $\hat{K}^{(0)}$ can be constructed by the Knonecker products of eigenstates for the subspace.
The left eigenstates are
\begin{align}\label{left_eigen_0} 
\langle \ell^{(0)}_0|&=\langle e|:=\langle 0| \otimes \langle 0|, \notag\\
\langle \ell^{(0)}_1|&=\langle\uparrow|:=\langle 1| \otimes \langle 0 |, \notag\\
\langle \ell^{(0)}_2|&=\langle\downarrow|:= \langle 0| \otimes \langle 1 |, \notag\\ 
\langle \ell^{(0)}_3|&=\langle d|:=\langle 1| \otimes \langle 1| ,
\end{align} 
and the right eigenstates ate
\begin{align}\label{right_eigen_0} 
|r^{(0)}_0\rangle&=|e\rangle:=|0\rangle \otimes |0\rangle, \notag\\
|r^{(0)}_1\rangle&=|\uparrow\rangle:=|1\rangle \otimes |0\rangle, \notag\\
|r^{(0)}_2\rangle&=|\downarrow\rangle:= |0\rangle \otimes |1\rangle, \notag\\ 
|r^{(0)}_3\rangle&=|d\rangle:=|1\rangle \otimes |1\rangle .
\end{align} 
These eigenstates satisfies orthonormal property, i. e. $\langle r^{(0)}_m|r^{(0)}_n\rangle=\delta_{mn}$.
The eigenvalues corresponding to $\langle \ell^{(0)}_m|$ and $|r^{(0)}_m\rangle$ are, respectively, given by
\begin{align}
\label{e_00}
\varepsilon_0^{(0)}&=2\mathcal{E}_0=0,\\
 \varepsilon_{1}^{(0)}&=\mathcal{E}_0+\mathcal{E}_1=-2, \label{e_01}\\
 \varepsilon_{2}^{(0)}&=\mathcal{E}_1+\mathcal{E}_0=-2 \label{e_10}\\
 \varepsilon_{3}^{(0)}&=2\mathcal{E}_1=-4,
 \label{e_11} 
\end{align} 
These eigenstates correspond to the empty ($m=0$), single occupied by the upspin ($m=1$), single occupied by the downspin ($m=2$) and double occupied ($m=3$) states, respectively.

\subsection{Perturbed states}

\subsubsection{$\hat{K}^{(1)}$ and basis for the degenerated system}

Similarly, $\hat{K}^{(1)}$ in Eq.~\eqref{K_1} can be rewritten as
\begin{align}
\hat{K}^{(1)}&=\mathring{g}_-
\begin{pmatrix}
-1 & -1 \\
1 & 1 \\
\end{pmatrix}
\otimes
\begin{pmatrix}
1 & 0 \\
0 & 0 \\
\end{pmatrix}
+\mathring{g}_-
\begin{pmatrix}
1 & 0 \\
0 & 0 \\
\end{pmatrix}
\otimes
\begin{pmatrix}
-1 & -1 \\
1 & 1 \\
\end{pmatrix}
,
\label{K^1_2}
\end{align}
where we have used $\mathring{g}_+=-\mathring{g}_-$ derived from $\partial (g_+^{(1)}+g_-^{(1)})/\partial \tilde{\beta}=0$ because of Eq.~\eqref{sum_rule}.

Before moving the detailed calculation, we rearrange the eigenbasis for the unperturbed states, because $|\uparrow \rangle$ and $|\downarrow \rangle$ are degenerated.
With the aid of Eqs.~\eqref{rho22}, \eqref{rho_1} and \eqref{right_eigen_0}, one can check that $\hat{K}^{(1)}$ satisfies
\begin{align}
\label{K^1r_0}
\hat{K}^{(1)}|r_0^{(0)}\rangle &
=\frac{\mathring{g}_-}{2}\left\{\begin{pmatrix}
-1 \\ 1 \\
\end{pmatrix}
\otimes
\begin{pmatrix}
g_+ \\ 0 \\
\end{pmatrix}
+\begin{pmatrix}
g_+ \\ 0 \\
\end{pmatrix}
\otimes
\begin{pmatrix}
-1 \\ 1 \\
\end{pmatrix}
\right\}
, \\
\label{K^1up}
\hat{K}^{(1)}|\uparrow\rangle &=
\frac{\mathring{g}_-}{2}
\begin{pmatrix}
1 \\ 0 \\
\end{pmatrix}
\otimes
\begin{pmatrix}
-1 \\ 1\\
\end{pmatrix}
,
\\
\label{K^1down}
\hat{K}^{(1)}|\downarrow\rangle &
=\frac{\mathring{g}_-}{2}
\begin{pmatrix}
-1 \\ 1\\
\end{pmatrix}
\otimes
\begin{pmatrix}
1 \\ 0 \\
\end{pmatrix} ,
\\
\hat{K}^{(1)}|r_3^{(0)}\rangle &=
\frac{\mathring{g}_-}{2}
\left\{
\begin{pmatrix}
0 \\ 0 \\
\end{pmatrix}
\otimes
\begin{pmatrix}
1 \\ 0\\
\end{pmatrix}
+
\begin{pmatrix}
1 \\ 0\\
\end{pmatrix}
\otimes
\begin{pmatrix}
0 \\ 0 \\
\end{pmatrix}
\right\}
 .
\label{K^1r_3}
\end{align}
Thus, we obtain
\begin{align}\label{down_K^1_up}
\langle \downarrow |\hat{K}^{(1)}|\uparrow \rangle &=-\mathring{g}_-,
\\
\label{up_K^1_up}
\langle \uparrow  |\hat{K}^{(1)}|\uparrow \rangle &
=0, \\
\label{down_K^1_down}
\langle \downarrow |\hat{K}^{(1)}|\downarrow \rangle &
=0,
\\
\langle \uparrow  |\hat{K}^{(1)}|\downarrow \rangle &
=-\mathring{g}_-.
\label{up_K^1_down}
\end{align}

Therefore, we choose the following basis:
\begin{align}\label{tilde_00}
\langle \tilde{\ell}^{(0)}_0|&=\langle \ell^{(0)}|, \quad |\tilde{r}^{(0)}_0\rangle=|r^{(0)}_0\rangle \\
\langle \tilde{\ell}^{(0)}_1|&=\langle \uparrow|+\langle \downarrow | ,
\quad
|\tilde{r}^{(0)}_1\rangle=\frac{1}{2}(|\uparrow\rangle+|\downarrow\rangle )
\label{tilde_10}
\\
\langle \tilde{\ell}^{(0)}_2|&=\langle \uparrow|-\langle \downarrow | ,
\quad
|\tilde{r}^{(0)}_2\rangle=\frac{1}{2}(|\uparrow\rangle-|\downarrow\rangle ) , 
\label{tilde_01}
\\ 
\langle\tilde{\ell}^{(0)}_3|&=\langle \ell^{(0)}_3|, \quad |\tilde{r}^{(0)}_3\rangle=|r^{(0)}_3\rangle .
\label{tilde_11} 
\end{align}
This set of basis also satisfies the orthonormal relation $\langle \tilde{\ell}^{(0)}_m|\tilde{r}^{(0)}_n\rangle=\delta_{mn}$.
The eigenvalues corresponding to $\langle\tilde{\ell}^{(0)}_m|$ and $|\tilde{r}^{(0)}_m\rangle$ are unchanged as
\begin{align}\label{zeroth_eigenvalues}
\tilde{\varepsilon}_0^{(0)}=0, {~}
\tilde{\varepsilon}_1^{(0)}=-2, {~}\tilde{\varepsilon}_2^{(0)}=-2, {~} \tilde{\varepsilon}_3^{(0)}=-4 .
\end{align}

It is easy to verify the completeness as
\begin{align} 
\sum_{i=0}^3 |\tilde{r}^{(0)}\rangle \langle \tilde{\ell}^{(0)}| 
&= |0\rangle \langle 0|\otimes |0\rangle \langle 0|+
|1\rangle \langle 1|\otimes |0\rangle \langle 0|
\notag\\
&\quad +
 |0\rangle \langle 0| \otimes |1\rangle \langle 1| +
 |1\rangle \langle 1|\otimes |1\rangle \langle 1| \notag \\
 &=\begin{pmatrix}
 1 & 0\\
 0 & 1\\
 \end{pmatrix}
 \otimes
 \begin{pmatrix}
 1 & 0\\
 0 & 1\\
 \end{pmatrix}.
\end{align}

\subsubsection{Perturbation analysis}

Now, let us obtain the perturbed eigenstates.
As the usual perturbation method, we expand the eigenstates and eigenvalues with the aid of $\langle \tilde{\ell}^{(0)}_m|$ and $|\tilde{r}^{(0)}_m\rangle$ as
\begin{align}\label{tilde_left_eigenvector}
\langle \ell_m | &=\langle \tilde{\ell}^{(0)}_m|+\tilde{\beta}\sum_{n=0}^3 a_{mn} \langle \tilde{\ell}^{(0)}_n|,
\\
|r_m\rangle &= |\tilde{r}^{(0)}_m\rangle +\tilde{\beta}\sum_{n=0}^3 b_{mn} |\tilde{r}^{(0)}_n\rangle, 
\label{tilde_right_eigenvector}
\\
\varepsilon_m &=\tilde{\varepsilon}^{(0)}_m+\tilde{\beta} \varepsilon^{(1)}_m.
\label{tilde_eigenvalue}
\end{align}   
Substituting Eqs.~\eqref{tilde_left_eigenvector}, \eqref{tilde_right_eigenvector} and \eqref{tilde_eigenvalue} into Eqs.~\eqref{left_eigen_eq} or \eqref{right_eigen_eq}
 multiplying $|\tilde{r}^{(0)}_k\rangle$ or $\langle \tilde{\ell}^{(0)}_k|$, we obtain
\begin{equation}\label{varepsilon_1}
\tilde{\varepsilon}^{(1)}_m=\langle \tilde{\ell}^{(0)}_m|\hat{K}^{(1)}|\tilde{r}_m^{(0)}\rangle .
\end{equation}
Subistituting Eqs.~\eqref{K^1r_0} and \eqref{K^1r_3} into Eq.~\eqref{varepsilon_1} we immediately obtain
\begin{equation}
\tilde{\varepsilon}_0^{(1)}=\tilde{\varepsilon}_3^{(1)}=0 ,
\end{equation}
due to $\langle \tilde{\ell}^{(0)}_0|\hat{K}^{(1)}|\tilde{r}^{(0)}_0\rangle=\langle \tilde{\ell}^{(0)}_3|\hat{K}^{(1)}|\tilde{r}^{(0)}_3\rangle=0$.
For $\tilde{\varepsilon}_1^{(1)}$ and $\tilde{\varepsilon}_2^{(1)}$, with the aid of Eqs.~(\ref{down_K^1_up},\ref{up_K^1_up},\ref{down_K^1_down},\ref{up_K^1_down}) and (\ref{tilde_10},\ref{tilde_01}) we obtain
\begin{align}
\tilde{\varepsilon}_1^{(1)}&=\frac{1}{2}(\langle \uparrow|+\langle \downarrow|)\hat{K}^{(1)}(|\uparrow\rangle+|\downarrow \rangle ) 
=-\mathring{g}_-, \\
\tilde{\varepsilon}_2^{(1)}&=\frac{1}{2}(\langle \uparrow|-\langle \downarrow|)\hat{K}^{(1)}(|\uparrow\rangle-|\downarrow \rangle ) 
=\mathring{g}_- .
\end{align}

Similarly, using Eqs.~\eqref{tilde_left_eigenvector}, \eqref{tilde_right_eigenvector}, \eqref{left_eigen_eq} and \eqref{right_eigen_eq} we obtain the relations at $O(\tilde{\beta})$ for $m\ne n$ as
\begin{align}\label{a_mn}
a_{mn}&=\frac{\langle \tilde{\ell}^{(0)}_m|\hat{K}^{(1)}|\tilde{r}_n^{(0)}\rangle}{\tilde{\varepsilon}_m^{(0)}-\tilde{\varepsilon}_n^{(0)}}=-b_{nm} ,\\
b_{mn}&=\frac{\langle \tilde{\ell}^{(0)}_n|\hat{K}^{(1)}|\tilde{r}_m^{(0)}\rangle}{\tilde{\varepsilon}_m^{(0)}-\tilde{\varepsilon}_n^{(0)}}.
\label{b_mn}
\end{align}
Then, we obtain
\begin{align}
\label{b01}
b_{01}&=-\frac{1}{2}\mathring{g}_-g_+ , \\
b_{02}&=b_{10}=b_{20}=b_{23}=b_{30}=b_{31}=b_{32}=0  ,\\
b_{03}&=-\mathring{g}_-g_+g_- ,
\label{b03}
\\
b_{13}&=-\frac{1}{2}\mathring{g}_-g_-
\label{b13}
\end{align}
We also have the relations
\begin{align}
\label{a10}
a_{10}&=\frac{1}{2}\mathring{g}_-g_+, \\
a_{20}&=a_{01}=a_{02}=a_{32}=a_{03}=a_{13}=a_{23}=0 ,\\
a_{30}&=\mathring{g}_-g_+g_-, \\
a_{31}&=\frac{1}{2}\mathring{g}_-g_- .
\label{a31}
\end{align}
Note that Eq.~\eqref{b_mn} cannot be used for $b_{12}$ and $b_{21}$ (as well as $a_{12}$ and $a_{21}$) because of the degeneracy of the ground state energy, i.e. zero denominator in Eq.~\eqref{b_mn}.
In this case, we may replace the denominator with $\tilde{\varepsilon}^{(1)}_1-\tilde{\varepsilon}^{(1)}_2$ for $b_{12}$.
If we accept this approximation, we estimate $b_{12}$, $b_{21}$, $a_{12}$ and $a_{21}$ as
\begin{align}\label{b12}
b_{12}&= b_{21}=a_{12}=a_{21}=0.
\end{align}
Substituting these relations into Eq.~\eqref{tilde_right_eigenvector} we obtain
\begin{align}
\label{pert_l0}
\langle \ell_0 |
&=\langle e|,  \\
|r_0\rangle &=|e\rangle -\tilde{\alpha} \left[
\frac{1}{2}g_+|\tilde{r}_1^{(0)}\rangle +g_+g_-|d\rangle  \right]
, \\
\langle \ell_1|&=\langle \tilde{\ell}_1^{(0)}|+\frac{1}{2}\tilde{\alpha}g_+\langle e| ,\\
|r_1\rangle &= |\tilde{r}_1^{(0)}\rangle-\frac{1}{2}\tilde{\alpha}g_-|d \rangle, \\
\langle \ell_2|&=\langle \tilde{\ell}_2^{(0)}|, \\
|r_2\rangle &=|\tilde{r}_2^{(0)}\rangle, \\
\langle \ell_3|&=\langle d|-\frac{\tilde{\alpha}}{2}
\left[
2g_+g_-\langle e|+g_-\langle \tilde{\ell}_1^{(0)}|
\right], 
\\
|r_3\rangle &=|d\rangle ,
\label{pert_r3}
\end{align}
where we have introduced $\tilde{\alpha}:=\tilde{\beta}\mathring{g}_-$.

\begin{widetext}
It is easy to verify that Eqs.~\eqref{pert_l0}-\eqref{pert_r3} satisfy the completeness relation as
\begin{align}
\sum_{m=0}^3|r_m\rangle\langle \ell_m|
&=\sum_{m=0}^3s
\{
|\tilde{r}_m^{(0)}\rangle+\tilde{\beta}\sum_{n=0}^3b_{mn}|\tilde{r}_n^{(0)}\rangle
\}\{
\langle \tilde{\ell}_m^{(0)}|+\tilde{\beta}\sum_{n=0}^3a_{mn}\langle \tilde{\ell}^{(0)}_n|
\} \notag\\
&=\sum_{m=0}^3|\tilde{r}_m^{(0)}\rangle \langle \tilde{\ell}^{(0)}_m|
+\tilde{\beta}\sum_{m,n}(b_{mn}+a_{nm})|\tilde{r}_n^{(0)}\rangle\langle\tilde{\ell}_m^{(0)}| +O(\tilde{\beta}^2) \notag\\
&=\sum_{m=0}^3|\tilde{r}_m^{(0)}\rangle \langle \tilde{\ell}^{(0)}_m|=1 ,
\end{align} 
where we have used Eq.~\eqref{a_mn}.


\subsection{Determination of $\hat{\rho}^{(1)}$ }

In this subsection, let us determine $\hat\rho^{(1)}$ in the super-vector notation based on 
\begin{equation}
|\hat\rho^{(1)}(\bm \Lambda(\theta))\rangle
=\dot\Lambda_\nu(\theta)\hat K^+(\bm \Lambda(\theta))\frac{\partial }{\partial \Lambda_\nu(\theta)}|\hat\rho^{\rm ss}(\bm \Lambda(\theta))\rangle,
\label{rho1}
\end{equation}
where $|\hat\rho^{\rm ss}(\bm \Lambda(\theta))\rangle$ is the eigenvector of $\hat K$ with the eigenvalue $0$, namely, $|r_0(\bm \Lambda(\theta))\rangle$.
In the present case, the super-vector $|r_0(\bm \Lambda(\theta))\rangle$ has 4-components which correspond to the diagonal entries of $\hat\rho^{\rm ss}(\bm \Lambda(\theta))$, thus the left hand side also has 4 components and which appear in the diagonal part in matrix representation. 
Substituting the spectrum expansion of the pseudo-inverse into Eq.\,(\ref{rho1}) yields 
\begin{align}
|\hat\rho^{(1)}(\bm \Lambda(\theta))\rangle
&=\dot\Lambda_\nu(\theta)
\sum_{m\neq 0}\frac{1}{\varepsilon_m(\bm \Lambda(\theta))}|r_m(\bm \Lambda(\theta))\rangle \langle \ell_m(\bm \Lambda(\theta))|
\frac{\partial }{\partial \Lambda_\nu(\theta)}|r_0(\bm \Lambda(\theta))\rangle\nonumber\\
&=\dot\Lambda_\nu(\theta)
\sum_{m\neq 0}\frac{1}{\varepsilon_m(\bm \Lambda(\theta))}|r_m(\bm \Lambda(\theta))\rangle 
\left[\frac{\partial }{\partial \Lambda_\nu(\theta)}\langle \ell_m(\bm \Lambda(\theta))|r_0(\bm \Lambda(\theta))\rangle
-\left(\frac{\partial }{\partial \Lambda_\nu(\theta)}\langle \ell_m(\bm \Lambda(\theta))|\right)|r_0(\bm \Lambda(\theta))\rangle\right]\nonumber\\
&=-\dot\Lambda_\nu(\theta)
\sum_{m\neq 0}\frac{1}{\varepsilon_m(\bm \Lambda(\theta))}|r_m(\bm \Lambda(\theta))\rangle 
\left(\frac{\partial }{\partial \Lambda_\nu(\theta)}\langle \ell_m(\bm \Lambda(\theta))|\right)|r_0(\bm \Lambda(\theta))\rangle.
\end{align}
Since the left eigenvector $\langle \ell_2|$ does not depend on $\bm \Lambda$ in the present case (see Eq.~\eqref{left_2}), it can be further simplified as 
\begin{align}\label{rho1_2}
|\hat\rho^{(1)}(\bm \Lambda(\theta))\rangle
&=-\dot\Lambda_\nu(\theta)
\sum_{m=1,3}\frac{1}{\varepsilon_m(\bm \Lambda(\theta))}|r_m(\bm \Lambda(\theta))\rangle 
\left(\frac{\partial }{\partial \Lambda_\nu(\theta)}\langle \ell_m(\bm \Lambda(\theta))|\right)|r_0(\bm \Lambda(\theta))\rangle.
\end{align}
Thus, once we obtain the eigenfuctions and eigenvalues of $\hat{K}$, we can determine $\hat{\rho}^{(1)}$.

\subsection{Entropy production and thermodynamic length}

In this subsection we briefly explain how to calculate the entropy production $\sigma^{(2)}$ and thermodynamic length $\mathcal{L}$.
For this purpose we use the expression in Eq.~\eqref{rho1_2} for the perturbative density matrix $\hat{\rho}^{(1)}$. 

By using eigenvalues, eigenstates, and formula for the $|\hat\rho^{(1)}\rangle$ obtained in the previous subsection, we calculate the one-cycle averaged entropy production and the thermodynamic length in this subsection. 
Substitution of Eqs.\,(\ref{left_1}), (\ref{right_1}), (\ref{left_3}), and (\ref{right_3}) into Eq.\,\eqref{rho1_2} yields  
\begin{align}\label{rho1exp}
&|\hat\rho^{(1)}\rangle=
\frac{2\dot\Lambda_\nu}{\left(f_+^{(0)}+f_-^{(1)}\right)^2}\left(
f_+^{(0)}\frac{\partial f_-^{(1)}}{\partial \Lambda_\nu}
-f_-^{(1)}\frac{\partial f_+^{(0)}}{\partial \Lambda_\nu}\right)|r_1\rangle\notag\\
&+\frac{\dot\Lambda_\nu}{4\left(f_+^{(0)}+f_-^{(1)}\right)}\left(
f_+^{(1)}f_-^{(1)} \frac{\partial f_+^{(0)}}{\partial \Lambda_\nu}
+f_+^{(0)}f_-^{(0)} \frac{\partial f_-^{(1)}}{\partial \Lambda_\nu}
\right)|r_3\rangle\notag\\
&=\frac{1}{8\left(f_+^{(0)}+f_-^{(1)}\right)^3\left(f_-^{(0)}+f_+^{(1)}\right)}\notag\\
&\hspace{1em}\times\left[8\left( f_+^{(0)}\frac{df_-^{(1)}}{d\theta}-f_-^{(1)}\frac{df_+^{(0)}}{d\theta}\right)
\begin{pmatrix}
2f_+^{(1)} \\[0.5em]
f_-^{(1)}-f_+^{(0)}\\[0.5em]
f_-^{(1)}-f_+^{(0)}\\[0.5em]
-2f_-^{(0)} 
\end{pmatrix}\right.\notag\\
&\hspace{2em}+\left.\left(f_+^{(0)}+f_-^{(1)}\right)^2
\left( f_+^{(1)} f_-^{(1)}\frac{df_+^{(0)}}{d\theta}+ f_+^{(0)} f_-^{(0)}\frac{df_-^{(1)}}{d\theta}\right)
\begin{pmatrix}
1 \\[0.5em]
-1\\[0.5em]
-1\\[0.5em]
1
\end{pmatrix}\right]
,
\end{align}
where we have used
\begin{align}\label{rho1exp2}
&\frac{\partial \langle \ell_1|}{\partial \Lambda_\nu}|r_0\rangle
=\frac{2}{f_+^{(0)}+f_-^{(1)}}
\left(
f_+^{(0)}\frac{\partial f_-^{(1)}}{\partial \Lambda_\nu}
-f_-^{(1)}\frac{\partial f_+^{(0)}}{\partial \Lambda_\nu}
\right),\\
&\frac{\partial \langle \ell_3|}{\partial \Lambda_\nu}|r_0\rangle
=
\frac{1}{f_+^{(0)}+f_-^{(1)}}\left(
f_+^{(1)}f_-^{(1)} \frac{\partial f_+^{(0)}}{\partial \Lambda_\nu}
+f_+^{(0)}f_-^{(0)} \frac{\partial f_-^{(1)}}{\partial \Lambda_\nu}
\right).
\end{align}
By rewriting Eq.\,(\ref{rho1exp2}) in the matrix form, and substituting it into Eq.~(\ref{A-ad-1}) we can calculate $\sigma^{(2)}$ in Eq.~\eqref{Eq(30)}.
Taking the square-root of the integrant of $\sigma^{(2)}$ and integrate Eq.~\eqref{L}, we can also obtain $\mathcal{L}$.


\subsection{Thermodynamic length}

By using eigenvalues and eigenstates obtained in the previous subsection, we calculate the thermodynamic length in this subsection. 
First we rewrite the eigenstates for the unperturbed part. 
The right eigenstates written as the Kronecker product of spin-up and spin-down subspace can be rewritten as follows 
\begin{align}
|e\rangle 
=|0\rangle\otimes |0\rangle
= 
\frac{1}{2}\left(
\begin{array}{c}
g_+\\
g_-
\end{array}
\right)\otimes 
\frac{1}{2}\left(
\begin{array}{c}
g_+\\
g_-
\end{array}
\right)
=
\frac{1}{4}\left(
\begin{array}{c}
\left(
\begin{array}{c}
g_+\\
g_-
\end{array}
\right)g_+\\
\left(
\begin{array}{c}
g_+\\
g_-
\end{array}
\right)g_-
\end{array}
\right)
=\frac{1}{4}\left(
\begin{array}{c}
g_+^2\\
g_+g_-\\
g_+g_-\\
g_-^2\\
\end{array}
\right).  
\end{align}
The same procedures yield 
\begin{align}
&|\tilde{r}_1^{(0)}\rangle 
=\frac{1}{2}\left(|\uparrow\rangle +|\downarrow\rangle \right)
=
\frac{1}{2}\left[
\frac{1}{4}\left(
\begin{array}{c}
g_+\\
g_-\\
-g_+\\
-g_-\\
\end{array}
\right)+
\frac{1}{4}\left(
\begin{array}{c}
g_+\\
-g_+\\
g_-\\
-g_-\\
\end{array}
\right)
\right]
=\frac{1}{4}\left(
\begin{array}{c}
g_+\\
1-g_+\\
1-g_+\\
-g_-\\
\end{array}
\right),\\
&|\tilde{r}_2^{(0)}\rangle 
=\frac{1}{2}\left(|\uparrow\rangle -|\downarrow\rangle \right)
=
\frac{1}{2}\left[
\frac{1}{4}\left(
\begin{array}{c}
g_+\\
g_-\\
-g_+\\
-g_-\\
\end{array}
\right)-
\frac{1}{4}\left(
\begin{array}{c}
g_+\\
-g_+\\
g_-\\
-g_-\\
\end{array}
\right)
\right]
=\frac{1}{4}\left(
\begin{array}{c}
0\\
1\\
-1\\
0\\
\end{array}
\right),\\
&|d\rangle 
=\frac{1}{4}\left(
\begin{array}{c}
1\\
-1\\
-1\\
1\\
\end{array}
\right).
\end{align}
For the left eigenstates we obtain 
\begin{align}
&\langle e|=(1,1,1,1),\\
&\langle \tilde{\ell}_1^{(0)}|=(g_-, g_-, -g_+,-g_+)+(g_-, -g_+, g_-,-g_+)=2(g_-,1-g_+,1-g_+,-g_+),\\
&\langle \tilde{\ell}_2^{(0)}|=(g_-, g_-, -g_+,-g_+)-(g_-, -g_+, g_-,-g_+)=2(0,1,-1,0),\\
&\langle d|=(g_-^2,-g_+g_-,-g_+g_-,g_+^2).
\end{align}
Thus we can rewrite the perturbed eigenstate $|r_0\rangle=|\hat{\rho}^{\rm ss}\rangle$ in Eq.~\eqref{right_zero} as 
\begin{align}
|r_0\rangle
&=\frac{1}{4}\left(
\begin{array}{c}
g_+^2\\
g_+g_-\\
g_+g_-\\
g_-^2\\
\end{array}
\right)
-\frac{1}{4}\tilde \alpha\left(
\begin{array}{c}
g_+\\
1-g_+\\
1-g_+\\
-g_-\\
\end{array}
\right)
-\frac{1}{8}\tilde \alpha g_+g_-
\left(
\begin{array}{c}
1\\
-1\\
-1\\
1\\
\end{array}
\right)\nonumber\\
&=\frac{1}{8}\left(
\begin{array}{c}
2g_+^2-2\tilde \alpha g_+-\tilde \alpha g_+g_-\\
2g_+g_- -2\tilde \alpha (1-g_+)+\tilde \alpha g_+g_-\\
2g_+g_- -2\tilde \alpha (1-g_+)+\tilde \alpha g_+g_-\\
2g_-^2-+2\tilde \alpha g_- -\tilde \alpha g_+g_-\\
\end{array}
\right).
\end{align}
Similarly, $|r_1\rangle$ obeys
\begin{align}
|r_1\rangle
&=\frac{1}{4}\left(
\begin{array}{c}
g_+\\
1-g_+\\
1-g_+\\
-g_-\\
\end{array}
\right)-\frac{1}{8}\tilde\alpha g_-\left(
\begin{array}{c}
1\\
-1\\
-1\\
1\\
\end{array}
\right)
=\frac{1}{8}\left(
\begin{array}{c}
2g_+-\tilde\alpha g_-\\\
2-2g_++\tilde\alpha g_-\\\
2-2g_++\tilde\alpha g_-\\\
-2g_--\tilde\alpha g_-\\\
\end{array}
\right).
\end{align}
A parallel procedure for the left super-eigenvectors leads to 
\begin{align}
\langle \ell_1|
=&\frac{1}{2}(4g_-+\tilde \alpha g_+,4-4g_++\tilde \alpha g_+,4-4g_++\tilde \alpha g_+,-4g_++\tilde \alpha g_+),\\
\langle \ell_3|=&(g_-^2-2\tilde\alpha g_-,-g_+g_--\tilde\alpha g_-,-g_+g_--\tilde\alpha g_-,g_+^2).
\end{align}
With the aid of the relation $g_++g_-=2$, one can show  
\begin{align}
\frac{\partial}{\partial \Lambda_\mu } \langle \ell_1|
=&\frac{1}{2}\frac{\partial (-4+\tilde\alpha )g_+}{\partial \Lambda_\mu}(1,1,1,1) .
\end{align}
Thus, we obtain 
\begin{align}
\left(\frac{\partial}{\partial \Lambda_\mu } \langle \ell_1|\right)|r_0\rangle 
=&\frac{1}{2}\frac{\partial (-4+\tilde\alpha )g_+}{\partial \Lambda_\mu}=-2\frac{\partial g_+}{\partial \Lambda_\mu}+O(\tilde\beta).
\label{l1r0}
\end{align}
Similarly, from the relation 
\begin{align}
\frac{\partial}{\partial \Lambda_\mu } \langle \ell_3|
=&\frac{\partial}{\partial \Lambda_\mu } (g_-^2,-g_+g_-,-g_+g_-, g_+^2)-\frac{\partial (\tilde \alpha g_-)}{\partial \Lambda_\mu } (2,1,1,0),
\end{align}
we obtain 
\begin{align}
\left(\frac{\partial}{\partial \Lambda_\mu } \langle \ell_3|\right)|r_0\rangle 
=& \left[\frac{1}{4}\frac{\partial}{\partial \Lambda_\mu } (g_-^2,-g_+g_-,-g_+g_-, g_+^2)\right]
\left(
\begin{array}{c}
g_+^2\\
g_+g_-\\
g_+g_-\\
g_-^2\\
\end{array}
\right)+O(\tilde \beta)\nonumber\\ 
=&\frac{1}{4}\left(\frac{\partial g_-^2}{\partial \Lambda_\mu } g_+^2-2\frac{\partial g_+g_-}{\partial \Lambda_\mu }g_+g_-+\frac{\partial g_+^2}{\partial \Lambda_\mu } g_-^2\right)
+O(\tilde \beta )=O(\tilde \beta).
\label{l3r0}
\end{align}

Substituting Eqs.\,(\ref{l1r0}) and (\ref{l3r0}) into Eq.\,(\ref{rho1_2}), the leading order of $|\rho^{(1)}\rangle$ in $\tilde\beta$ expansion becomes 
\begin{align}
|\hat\rho^{(1)}(\bm \Lambda(\theta))\rangle
&=2\dot\Lambda_\nu(\theta)\frac{\partial g_+}{\partial \Lambda_\nu}\frac{1}{\varepsilon_1(\bm \Lambda(\theta))}|r_1(\bm \Lambda(\theta))\rangle+O(\tilde{\beta})
\nonumber\\
&=2\dot\Lambda_\nu(\theta)\frac{\partial g_+}{\partial \Lambda_\nu}\frac{1}{\tilde \varepsilon_1(\bm \Lambda(\theta))}|\tilde r_1(\bm \Lambda(\theta))\rangle+O(\tilde{\beta})
\nonumber\\
&=-\frac{d g_+}{d\theta}|\tilde r_1(\bm \Lambda(\theta))\rangle+O(\tilde{\beta}).
\label{rho1leading}
\end{align}
Thus, we can rewrite Eq.\,(\ref{rho1leading}) in the matrix form as 
\begin{align}
\hat\rho^{(1)}(\bm \Lambda(\theta))
=-\frac{1}{4}\frac{d g_+}{d\theta}\mathrm{diag} (g_+, 1-g_+, 1-g_+, -g_-)+O(\tilde{\beta}).
\label{rho1leading}
\end{align}
Finally, we get 
\begin{align}
\sigma^{(2)}&=\frac{1}{4\pi}\int_0^{2\pi} d\theta\frac{1}{4}\left(\frac{d g_+}{d\theta}\right)^2
\mathrm{Tr}\{ \mathrm{diag} [1, (1-g_+)^2/g_+g_-, (1-g_+)^2/g_+g_-, 1]\}+O(\tilde{\beta})
\nonumber\\
&=\frac{1}{4\pi}\int_0^{2\pi} d\theta\frac{1}{2}\left(\frac{d g_+}{d\theta}\right)^2
\frac{g_+g_-+(1-g_+)^2}{g_+g_-}+O(\tilde{\beta}) \nonumber\\
&=\frac{1}{8\pi}\int_0^{2\pi} d\theta\left(\frac{d g_+}{d\theta}\right)^2
\frac{1}{g_+g_-}+O(\tilde{\beta}).
\end{align}
The integrant of the leading term is positive definite as expected. 
Similar calculation yields 
\begin{align}\label{thermo_length}
\mathcal{L}&= \frac{1}{4\pi}\int_0^{2\pi} d\theta\sqrt{\left(\frac{d g_+}{d\theta}\right)^2\frac{1}{g_+g_-}}+O(\tilde{\beta})
= \frac{1}{4\pi}\int_0^{2\pi} d\theta \left|\frac{d g_+}{d\theta}\right|\frac{1}{\sqrt{g_+g_-}}+O(\tilde{\beta}).
\end{align}


\subsection{Adiabatic work}


In this subsection, we present the details of the adiabatic work $\mathcal{W}$. 

By using the curvature introduced in Eq.~\eqref{Gdef}, the 2-form $d\mathcal{A}$ can be written as $d\mathcal{A}=\frac{1}{2}F_{\mu\nu}d\Lambda_\mu\wedge d\Lambda_\nu$. 
If both of $\Lambda_\mu$ and $\Lambda_\nu$ are not $\lambda$, $A_\nu$ for $\partial \hat{H}/\partial\lambda=0$ at $\theta=2\pi$ satisfies the relation
\begin{equation}
\frac{\partial}{\partial \Lambda_\mu} A_\nu
=-\frac{\partial}{\partial \Lambda_\mu} \left[\lambda \frac{\partial}{\partial \Lambda_\nu}\mathrm{Tr}\left(\frac{\partial\hat H}{\partial\lambda}\hat \rho^{\rm ss}\right)\right]
= -\lambda \frac{\partial}{\partial \Lambda^\mu}\frac{\partial}{\partial \Lambda_\nu}\mathrm{Tr}\left(\frac{\partial\hat H}{\partial\lambda}\hat \rho^{\rm ss}\right)
=\frac{\partial}{\partial \Lambda_\nu} A_\mu .
\end{equation}
Thus, $F_{\mu\nu}$ can be nonzero only if $\Lambda_\mu$ or $\Lambda_\nu$ is chosen to be $\lambda$. 
For instance, if we choose $\mu=3, \nu=1$ with $\Lambda_3=\lambda$ and $\Lambda_1=\mu^L/\bar \mu^{\rm L}$, we obtain $F_{31}$ as
\begin{equation}
F_{31}=-\overline{\mu} \frac{\partial}{\partial \mu^L}\mathrm{Tr}\left(\frac{\partial\hat H}{\partial\lambda}\hat \rho^{\rm ss}\right). 
\label{curvature01}
\end{equation}

If the Hamiltonian $\hat H$ depends on $\lambda$ only through $U=U_0\lambda(\theta)$ as we consider in this paper, one can further calculate the thermodynamic curvature (\ref{curvature01}) as 
\begin{align}
F_{31}&=-\overline{\mu} \frac{\partial}{\partial \mu^{\rm L}}\mathrm{Tr}\left(U_0 \hat{n}_{\uparrow}\hat{n}_{\downarrow}\hat \rho^{\rm ss}\right)
=-\overline{\mu} U_0 \frac{\partial}{\partial \mu^{\rm L}}( \rho^{\rm ss}_{11}(\mathbf{\Lambda}))\nonumber\\
&=-\frac{1}{8}\overline{\mu} U_0\frac{\partial}{\partial \mu^{\rm L}}(2g_+^2-2\tilde \alpha g_+-\tilde \alpha g_+g_-).
\end{align}
where we have used the matrix representation of the double occupied state $(\hat{n}_\uparrow \hat{n}_\downarrow)_{ij}=\delta_{i,1}\delta_{j,1}$. 
In order to visualize the curvature, we introduce the thermodynamic axial field $\vec{\mathscr{B}}$ as in Eq.~\eqref{B=rotA}. 

In the present case, the thermodynamic axial field $\vec{\mathscr{B}}$ is given by 
\begin{align}
\vec{\mathscr{B}}&=-
\frac{1}{8} \overline{\mu} U_0
\begin{pmatrix}
 \frac{\partial}{\partial \mu^{\rm L}}(2g_+^2-2\tilde \alpha g_+-\tilde \alpha g_+g_-)
\\
-\frac{\partial}{\partial \mu^{\rm R}}(2g_+^2-2\tilde \alpha g_+-\tilde \alpha g_+g_-)
\\ 
0\\
\end{pmatrix},
\end{align}
The axial field $\vec{\mathscr{B}}$ can be expanded as $\vec{\mathscr{B}}=\tilde{\beta}\vec{\mathscr{B}}^{(1)}$ for small $\tilde{\beta}=\bar \beta^{\rm L} U_0$ and $\tilde{\alpha}=\tilde{\beta}\mathring{f}_-$, where $\vec{\mathscr{B}}^{(1)}$ is given by
\begin{align}
\vec{\mathscr{B}}^{(1)}&=-\frac{1}{2}\overline{\mu}
\begin{pmatrix}
g_+\frac{\partial g_+}{\partial \mu^{\rm L}}
\\
- g_+\frac{\partial g_+}{\partial \mu^{\rm R}}  
\\
 0
\\
\end{pmatrix}
.
\end{align}
Thus, the thermodynamic axial field $\mathscr{B}_\mu$ is independent of $\lambda(\theta)$.
Therefore, the path in the parameter space $\Lambda_\mu$ is embedded in the two-dimensional space spanned by $\mu^{\rm L}$ and $\mu^{\rm R}$.

For a divergence free variable $\mathscr{B}_\mu$ satisfying Eq.~\eqref{divergence_free}, we can introduce the stream function $\Psi(\mu^{\rm L},\mu^{\rm R})$ satisfying 
\begin{align}
\vec{\mathscr{B}}
=\begin{pmatrix}
-\frac{\partial}{\partial \mu^{\rm L}} \Psi(\mu^{\rm L},\mu^{\rm R})
\\
\frac{\partial}{\partial \mu^{\rm R}}\Psi(\mu^{\rm L},\mu^{\rm R}) 
 \\
0\\
\end{pmatrix}
.
\end{align}
Note that $\Psi(\mu^{\rm L},\mu^{\rm R})$ is symmetric satisfying $\Psi(\mu^{\rm L},\mu^{\rm R})=\Psi(\mu^{\rm R},\mu^{\rm L})$.
Since the operational path enclosed by $\partial \Omega$ embedded in the two-dimensional space, 
the normal vector perpendicular to the enclosed path is proportional to $(1,1,0)^\tau$ where $\tau$ expresses the transverse for the case of $\delta=\pi$.
Thus, the thermodynamic flux penetrating the path $\partial \Omega$ at $\mu^{\rm L}=a$ and $\mu^{\rm R}=b$ is expressed as
\begin{equation}
\Phi_{\rm TD}(a,b)
\propto \vec{\mathscr{B}}\cdot 
\begin{pmatrix} 1\\ 1 \\ 0 \\ \end{pmatrix}
=
\frac{\partial}{\partial \mu^{\rm L}}\Psi(\mu^{\rm L},\mu^{\rm R})|_{\mu^{\rm L}=a,\mu^{\rm R}=b}
-
\frac{\partial}{\partial \mu^{\rm R}} \Psi(\mu^{\rm L},\mu^{\rm R})|_{\mu^{\rm L}=a,\mu^{\rm R}=b} .
\end{equation}
Since $\Psi(\mu^{\rm L},\mu^{\rm R})=\Psi(\mu^{\rm R},\mu^{\rm L})$, the thermodynamic flux satisfies
\begin{align}
\Phi_{\rm TD}(b,a)
&\propto
\frac{\partial}{\partial \mu^{\rm L}}\Psi(\mu^{\rm L},\mu^{\rm R})|_{\mu^{\rm L}=b,\mu^{\rm R}=a}
-
\frac{\partial}{\partial \mu^{\rm R}} \Psi(\mu^{\rm L},\mu^{\rm R})|_{\mu^{\rm L}=b,\mu^{\rm R}=a} \notag\\
&= \frac{\partial}{\partial \mu^{\rm L}}\Psi(\mu^{\rm R},\mu^{\rm L})|_{\mu^{\rm L}=b,\mu^{\rm R}=a}
-
\frac{\partial}{\partial \mu^{\rm R}} \Psi(\mu^{\rm R},\mu^{\rm L})|_{\mu^{\rm L}=b,\mu^{\rm R}=a} \notag\\
&=-\frac{\partial}{\partial \mu^{\rm R}} \Psi(\mu^{\rm R},\mu^{\rm L})|_{\mu^{\rm R}=a,\mu^{\rm L}=b}
+
\frac{\partial}{\partial \mu^{\rm L}}\Psi(\mu^{\rm R},\mu^{\rm L})|_{\mu^{\rm R}=a,\mu^{\rm L}=b}.
\end{align}
Thus, one can find the relation $\Phi_{\rm TD}(b,a)=-\Phi_{\rm TD}(a,b)$.
This means that the thermodynamic flux penetrating the area becomes zero if the shape of the area is symmetric under the exchange of $\mu^{\rm L}$ and $\mu^{\rm R}$.
Thus, the adiabatic work $\mathcal{W}$ becomes zero at $\delta=\pi$, if the modulation of the parameter is cyclic.

Now, let us write the vector potential explicitly as
\begin{align}
\mathcal{A}_\mu&=\frac{1}{8}U_0\lambda \frac{\partial}{\partial \Lambda_\mu}
(2g_+^2-2\tilde{\alpha}g_+-\tilde{\alpha}g_+g_-) \notag\\
&=\frac{1}{8}U_0\lambda \frac{\partial}{\partial \Lambda_\mu}
(2g_+^2-2\beta  U_0 \mathring{g}_- g_+ -\beta  U_0\mathring{g}_- g_+g_-).
\end{align}
One can rewrite $\mathcal{A}_\mu$ furthermore
\begin{align}
\mathcal{A}_\mu
&=-\frac{1}{8} \frac{\partial}{\partial \Lambda_\mu}
\left[\lambda(2g_+^2-2 \beta  U_0 \mathring{g}_- g_+ -\beta  U_0\mathring{g}_- g_+g_-)\right]
+\frac{1}{8}\frac{\partial \lambda}{\partial \Lambda_\mu}(2g_+^2-2\beta U_0 \mathring{g}_- g_+ -\beta U_0\mathring{g}_- g_+g_-)
\notag\\
&=-\frac{1}{8} \frac{\partial}{\partial \Lambda_\mu}
\left[\lambda(2g_+^2-2\beta  U_0 \mathring{g}_- g_+ -\beta U_0\mathring{g}_- g_+g_-)\right]
+\frac{1}{8} U_0\delta_{\mu,3}(2g_+^2-2 \beta  U_0 \mathring{g}_- g_+ -
\beta  U_0\mathring{g}_- g_+g_-).
\label{Yoshii19_0921}
\end{align}
If the modulation of the parameters is carried out in $(\mu^{\rm L}, \mu^{\rm R})$ plane, with keeping $\lambda=$const., the adiabatic work vanishes as
\begin{equation}
\mathcal{W}_{(\mu^{\rm L},\mu^{\rm R})}
=-\frac{1}{8\pi}U_0 \lambda 
\left\{ 
\oint d\mu^{\rm L}\frac{\partial g_+^2}{\partial \mu^{\rm L}}+
\oint d\mu^{\rm R}\frac{\partial g_+^2}{\partial \mu^{\rm R}}
\right\}
 =0.
\end{equation} 
Thus, with the aid of Eq.~\eqref{Yoshii19_0921}, the modulation with changing $\lambda$ leads to the adiabatic work as
\begin{align}
\mathcal{W}=\frac{1}{2\pi}\oint d\Lambda_\mu \mathcal{A}_\mu
=\frac{1}{16\pi}U_0\oint d\lambda (2g_+^2-2\beta  U_0 \mathring{g}_- g_+ -\beta  U_0\mathring{g}_- g_+g_-)
\approx
\frac{1}{8\pi}U_0\oint d\lambda g_+^2,
\end{align}
where we have used the assumption of the perturbation i.e. $\beta U_0 \ll 1$ to obtain the final expression.

\end{widetext}



\end{document}